\documentclass[twocolumn,superscriptaddress,english,prl,showpacs,longbibliography]{revtex4-2}

\usepackage[normalem]{ulem}

\usepackage{listings}
\usepackage{xcolor}
\usepackage[colorlinks=true,urlcolor=blue,citecolor=blue,linkcolor=blue]{hyperref}
\usepackage{makecell}
\usepackage{amsmath}
\usepackage{enumitem} 
\usepackage{graphicx}
\usepackage{dcolumn}
\usepackage{bm}
\usepackage{algorithm}
\usepackage{algorithmic}
\usepackage{color}
\usepackage{array}
\usepackage{tabularx}
\usepackage{longtable}
\usepackage{amssymb}
\usepackage{graphicx}
\usepackage{color}
\usepackage{mathrsfs}
\usepackage{float}
\usepackage{indentfirst}
\usepackage{txfonts}
\usepackage{booktabs} 

\usepackage{geometry}
\usepackage{multirow}

\newcommand{\beq}{\begin{equation}}
\newcommand{\eeq}{\end{equation}}

\newcommand{\pmf}{P_{\mathrm{MF}}}
\newcolumntype{"}{@{\vrule width 0.8pt}}
\newcolumntype{'}{@{\vrule width 0.5pt}}

\renewcommand{\arraystretch}{1.2}

\begin{document}

\title{Free-Energy Machine for Combinatorial Optimization}




\author{Zi-Song Shen}
\thanks{These three authors contributed equally}
\affiliation{
 CAS Key Laboratory for Theoretical Physics, Institute of Theoretical Physics, Chinese Academy of Sciences, Beijing 100190, China
}
\affiliation{
 School of Physical Sciences, University of Chinese Academy of Sciences, Beijing 100049, China
}

\author{Feng Pan}
\thanks{These three authors contributed equally}
\affiliation{
 CAS Key Laboratory for Theoretical Physics, Institute of Theoretical Physics, Chinese Academy of Sciences, Beijing 100190, China
}

\author{Yao Wang}
\thanks{These three authors contributed equally}
 \affiliation{2012 lab, Huawei Technologies Co., Ltd., Shenzhen 518129, China}
 
\author{Yi-Ding Men}
\affiliation{
 School of Physical Sciences, University of Chinese Academy of Sciences, Beijing 100049, China
}
\affiliation{
School of Fundamental Physics and Mathematical Sciences, Hangzhou Institute for Advanced Study, UCAS, Hangzhou 310024, China
}
\author{Wen-Biao Xu}
\affiliation{
 School of Physical Sciences, University of Chinese Academy of Sciences, Beijing 100049, China
}
\affiliation{
School of Fundamental Physics and Mathematical Sciences, Hangzhou Institute for Advanced Study, UCAS, Hangzhou 310024, China
}
\author{Man-Hong Yung}
\affiliation{2012 lab, Huawei Technologies Co., Ltd., Shenzhen 518129, China}

\author{Pan Zhang}
\email{panzhang@itp.ac.cn}
\affiliation{
 CAS Key Laboratory for Theoretical Physics, Institute of Theoretical Physics, Chinese Academy of Sciences, Beijing 100190, China
}
\affiliation{
School of Fundamental Physics and Mathematical Sciences, Hangzhou Institute for Advanced Study, UCAS, Hangzhou 310024, China
}

\date{\today}

\begin{abstract}
Finding optimal solutions to combinatorial optimization problems is pivotal in both scientific and technological domains, within academic research and industrial applications. A considerable amount of effort has been invested in the development of accelerated methods that leverage sophisticated models and harness the power of advanced computational hardware. Despite the advancements, a critical challenge persists, the dual demand for both high efficiency and broad generality in solving problems. In this work, we propose a general method, Free-Energy Machine (FEM), based on the ideas of free-energy minimization in statistical physics, combined with automatic differentiation and gradient-based optimization in machine learning. The algorithm is flexible, solving various combinatorial optimization problems using a unified framework, and is efficient, naturally utilizing massive parallel computational devices such as graph processing units (GPUs) and field-programmable gate arrays (FPGAs). We benchmark our algorithm on various problems including the maximum cut problems, balanced minimum cut problems, and maximum $k$-satisfiability problems, scaled to millions of variables, across both synthetic, real-world, and competition problem instances. The findings indicate that our algorithm not only exhibits exceptional speed but also surpasses the performance of state-of-the-art algorithms tailored for individual problems. 
This highlights that the interdisciplinary fusion of statistical physics and machine learning opens the door to delivering cutting-edge methodologies that will have broad implications across various scientific and industrial landscapes.
\end{abstract}

\maketitle


\section{\label{sec:intro}Introduction}
Combinatorial optimization problems (COPs) are prevalent across a broad spectrum of fields, from science to industry, encompassing disciplines such as statistical physics, operations research, and artificial intelligence, among many others~\cite{du1998handbook}. 
However, most of these problems are non-deterministic polynomial-time-hard (NP-hard)~\cite{arora2009computational}, posing significant computational challenges. 
It is widely believed that exact algorithms are unlikely to provide efficient solutions unless NP=P. 
Consequently, a plethora of classical algorithms, including simulated annealing~\cite{kirkpatrick1983optimization} and various local search algorithms~\cite{selman1994noise,glover1998tabu,boettcher2001optimization}, have been devised and widely adopted for approximately solving the problem in practical settings.
It is worth noting that most of them realize series computations and were designed for CPUs. While some special problems with certain structures can be solved efficiently~\cite{barahona1982computational}, most of the practical hard problems remain intractable by the standard tools. 

In recent years, due to the remarkable emergence of massively parallel computational power given by Graphics Processing Units (GPUs) and Field-Programmable Gate Arrays (FPGAs), there has been a growing expectation for novel approaches. Many novel approaches have been developed. 
Notable examples include the Simulated Coherent Ising Machine (SimCIM)~\cite{tiunov2019annealing}, Noise Mean Field Annealing (NMFA)~\cite{king2018emulating} and the Simulated Bifurcation Machines (SBM)~\cite{goto2019combinatorial,goto2021high}, etc. They are originally inspired by simulating the mean-field dynamics of programmable specialized hardware devices called Ising machines~\cite{johnson2011quantum,inagaki2016large,honjo2021100,pierangeli2019large,mallick2020using,cai2020power,aadit2022massively,mohseni2022ising}, 
and have been shown to achieve even higher performance compared to the hardware version~\cite{king2018emulating, goto2021high}.
In addition to their high accuracy, a significant signature of the algorithms is their ability to simultaneously update variables which enables effective acceleration to address large-scale problems using GPUs and FPGAs~\cite{goto2019combinatorial,goto2021high}. 
However, these algorithms have their applicability predominantly limited to quadratic unconstrained binary optimization (QUBO) problems~\cite{kochenberger2014unconstrained}, or Ising formulations~\cite{lucas2014ising}.
This limitation becomes evident when addressing COPs inherently characterized by non-quadratic (e.g., higher-order $p$-spin glasses~\cite{gardner1985spin} for Boolean $k
$-satisfiability ($k$-SAT) problems~\cite{karp2010reducibility}) or non-binary features (e.g., the Potts glasses~\cite{wu1982potts}, coloring problems~\cite{jensen2011graph}, and community detections~\cite{newman2006modularity}).
To adapt these more complex problem structures to Ising formulations, additional conversion steps and significant overhead are required. They complicate the optimization problem and make them more challenging to solve when compared to the original formulations.


In this work, we propose to address the demands of combinatorial optimization in terms of generality, performance, and speed, drawing inspiration from 
statistical physics. Specifically, our approach is distinguished by its ability to solve general COPs without relying on Ising formulations, setting it apart from existing methods. From the statistical physics perspective, the cost function of a COP plays the rule of energy (also the free energy at zero temperature) of the spin glass system. Solving COPs amounts to finding configurations minimizing the energy~\cite{papadimitriou1998combinatorial, lucas2014ising}. 
However, directly searching for configurations minimizing the energy is difficult, because the landscape of energy is rugged, and searching may frequently be trapped by local minima of energy. In spin glass theory of statistical physics, this picture is described by the theory of replica symmetry breaking, which uses the organization of fixed points of mean-field solutions to characterize the feature of the rugged landscapes~\cite{mezard2002analytic}. 

\begin{figure*}[htbp]
\centering
\includegraphics[width=1\linewidth]{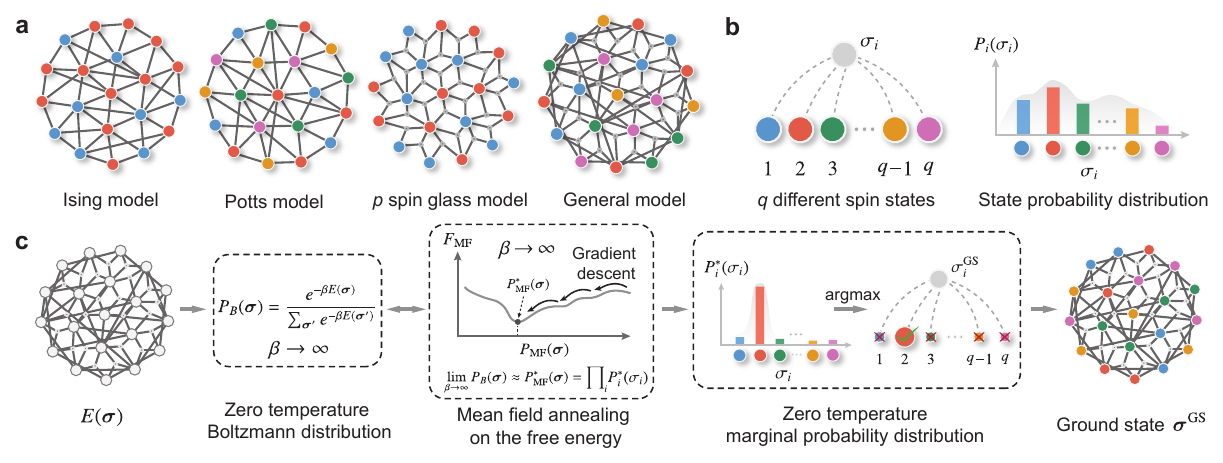}
\caption{\textbf{Illustration of the Free-Energy Machine in solving combinatorial optimization problems.}  \textbf{(a)}. Four distinct models offering representations of different types of combinatorial optimization problems. 
\textbf{(b)}. In a general combinatorial optimization problem (COP), each spin variable is capable of adopting one of $q$ distinct spin states, which are denoted as $1, 2, \ldots, q$. The likelihood of assuming any given spin state is described by the marginal probability $P_i(\sigma_i)$.
\textbf{(c)}. 
To tackle a combinatorial optimization problem (COP) characterized by the cost function $E(\bm{\sigma})$, the primary computational challenge involves calculating the marginal probability from the zero-temperature Boltzmann distribution. This calculation can be effectively approximated through a gradient-based annealing approach applied to the variational mean-field free energy $F_{\text{MF}}$. By determining the optimal mean-field distribution $P^*_{\text{MF}}(\bm{\sigma})$ that minimizes $F_{\text{MF}}$, it becomes possible to ascertain the ground state of the COP. This is achieved by identifying the most probable spin state for each spin variable, utilizing the set of marginal probabilities $P^*_i(\sigma_i)$.}
\label{fig:BolzmannDistribution}
\end{figure*}

Here, inspired by replica symmetry breaking,  we propose a general method based on minimizing variational free energies at a temperature that gradually annealed from a high value to zero. 
The free energies are functions of replicas of variational mean-field distributions and are minimized using gradient-based optimizers in machine learning.
We refer to our method as \textit{Free-Energy Machine}, abbreviated as FEM.
The approach incorporates two major features. First, the gradients of replicas of free energies are computed via automatic differentiation in machine learning, making it generic and immediately applied to various COPs. Second, the variational free energies are minimized by utilizing recognized optimization techniques such as Adam~\cite{kingma2014adam} developed from the deep learning community. Significantly, all replicas of mean-field probabilities are updated in parallel, thereby leveraging the computational power of GPUs for efficient execution and facilitating a substantial speed-up in solving large-scale problems. The pictorial illustration of our algorithm is shown in Fig.~\ref{fig:BolzmannDistribution}. 

We have evaluated FEM using a wide spectrum of combinatorial optimization challenges, each with unique features. This includes tackling the maximum cut (MaxCut) problem, fundamentally represented by the two-state Ising spin glasses; addressing the $q$-way balanced minimum cut (bMinCut) problem, which aligns with the Potts glasses and encapsulates COPs involving more than two states; and solving the maximum $k$-satisfiablity (Max $k$-SAT) problem, indicative of problems characterized by multi-body interactions. We measured FEM's efficacy by comparing it with the leading algorithms tailored to each specific problem. The comparative analysis reveals that the proposed approach not only competes well but in many instances outperforms these specialized, cutting-edge solvers across the board. This demonstrates FEM's exceptional adaptability and superior performance, both in terms of accuracy and efficiency, across a diverse set of COPs.

\begin{figure*}[htbp]
\centering
\includegraphics[width=1\linewidth]{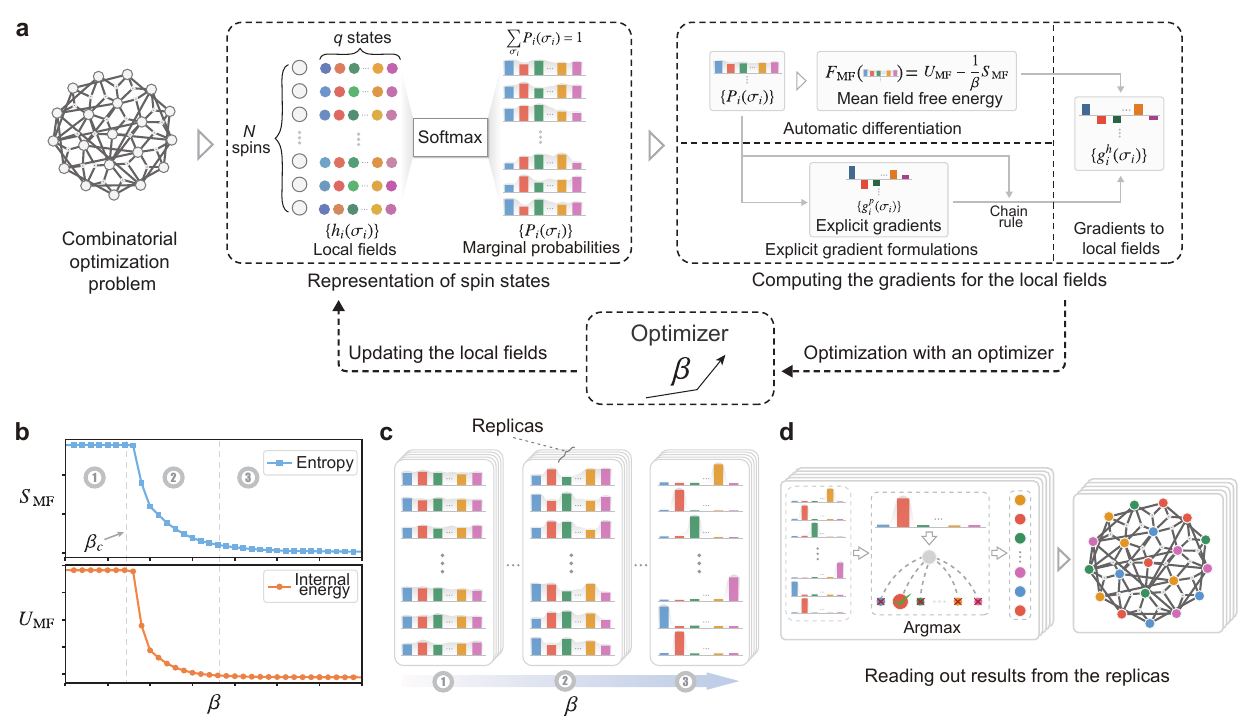}
\caption{\textbf{The framework of implementing Free-Energy Machine.}  \textbf{(a)}. Given a COP with $N$ spin variables, the spin states are associated with $N \times q$ variables called the local fields $\{h_i(\sigma_i)\}$. The marginal probabilities for all spin states are calculated from the local fields through the softmax function, and the local fields serve as the genuine variational variables, consistently guaranteeing the probabilistic interpretation of $\{P_i(\sigma_i)\}$ ($\sum_{\sigma_i}P_i(\sigma_i)$=1).  The gradients of mean-field free energy $F_{\rm MF}$ with respect to the local fields, denoted as $\{g_i^h(\sigma_i)\}$, can be computed via the automatic differentiation or through the explicit gradient formulations. In the explicit gradient formulations, $\{g_i^h(\sigma_i)\}$ can be computed via the chain rule using the explicit gradients $\{g_i^p(\sigma_i)\}$ (please refer to Supplementary Materials for details). The well-developed continuous optimizers from the deep learning community can be employed for the optimization. With the annealing, the local fields are updated to optimize $F_{\rm MF}$. \textbf{(b)}. The schematic of $U_{\rm MF}$ and $S_{\rm MF}$ evolving with the inverse temperature $\beta$ during the optimization process. $\beta_c$ is the critical inverse temperature at which the distribution transition occurs. \textbf{(c)}. The corresponding evolutions of the marginal probabilities with the annealing, depicting at the three stages marked in (b). Inspired by the one-step replica-symmetry breaking in statistical physics~\cite{mezard2002analytic}, we can update the marginal probabilities of many replicas parallelly.  \textbf{(d)}. Reading out the optimal solution from the replicas of marginal probabilities at the end of annealing. }
\label{fig:FEM}
\end{figure*}

\begin{figure*}[htbp]
\centering
\includegraphics[width=1\linewidth]{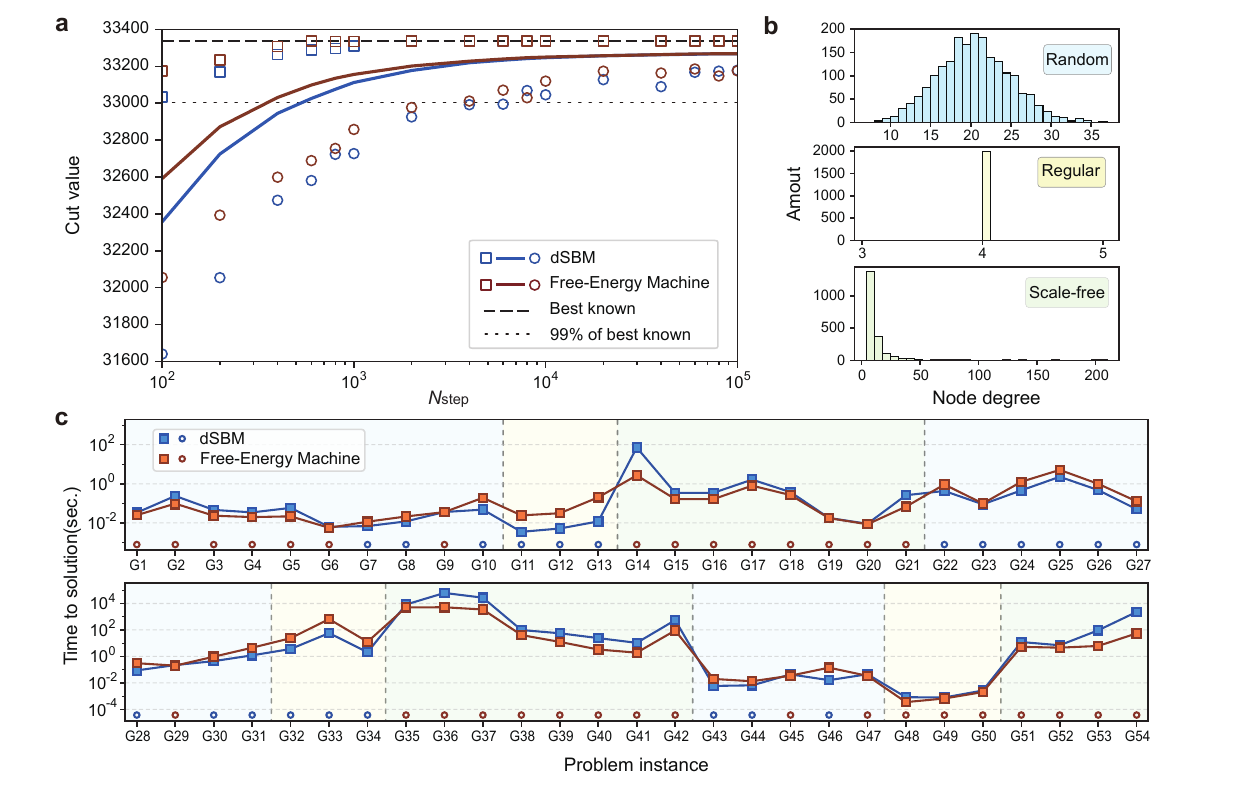}
\caption{\textbf{Benchmarking results for the maximum cut problem.}  \textbf{(a)}. The benchmarking results for solving the MaxCut problem on the complete graph $K_{\rm{2000}}$ with $2000$ nodes.
The graphical representations include squares, solid lines, and circles, which respectively illustrate the maximum, average, and minimum cut values as a function of the total number of annealing step $N_{\rm step}$. 
For each $N_{\rm step}$, we implement $R=1000$ mean-field replicas using FEM, showcasing the distribution of cut values derived from these replicas. In parallel, the discrete Stochastic Bifurcation Machine (dSBM) was executed for $1000$ trials with random initial conditions for each $N_{\text{step}}$ to serve as a comparative benchmark. 
The best-known cut value for $K_{\rm 2000}$ is 33337, with the 99\% of the best-known cut values approximately reaching 33004. An inverse-proportional annealing scheduling with $\beta$ ($T_{\rm max}=1.16$, $T_{\rm min}=6\times 10^{-5}$) was applied. Please refer to main text and Supplementary Materials for more details.
\textbf{(b)}.
The G-set instances commonly exhibit node degree distributions that categorize into three distinct types of graphs.
\textbf{(c)}. 
Time-to-Solution (TTS) benchmarking comparison of FEM and dSBM across G-set instances, which range from G1 to G54 and include graphs with 800 to 2000 nodes. 
The data for dSBM is referenced from the study in~\cite{goto2021high}. The G-set instances are visually categorized in the plot by regions marked with different colors, each color representing one of the three distinct graph types identified in these instances. Graph instances that exhibit shorter TTSs are notably highlighted with circles positioned at the lower section of the plot, indicating superior performance in those cases. For additional insights into the methodologies employed and the experimental setup, please refer to main text and Supplementary Materials.}
\label{fig:maxcut}
\end{figure*}

\section{Results}

\subsection{Free-Energy Machine}
\noindent Consider a COP characterized by a cost function, i.e. the energy function in physics, $E(\bm{\sigma})$, that we aim to minimize. 
Here, $\bm{\sigma}=(\sigma_1,\sigma_2,...,\sigma_N)$  represents a candidate solution or a configuration comprising $N$ discrete variables. 
The energy function encapsulates the interactions among variables, capturing the essence of various COPs. This is depicted in Fig.~\ref{fig:BolzmannDistribution}(a), where it is further delineated into four distinct models, each representing different physical scenarios.
\begin{enumerate}[leftmargin=10pt,labelindent=*, labelwidth=*,itemsep=0pt]
\item One of the simplest cases is the QUBO problem (or the Ising problem)~\cite{kochenberger2014unconstrained}, with $E(\bm{\sigma})=-\sum_{i<j}W_{ij}\sigma_i\sigma_j$, where $\sigma_i \in \{-1,+1\}$. 
The existing Ising solvers are tailor-made to address problems confined to the Ising model category. 
\item The COPs permitting variables to take multi-valued states, specifically $\sigma_i \in  \{1, 2, \ldots, q\}$, yet maintaining the two-body interactions, are categorized under the Potts model~\cite{wu1982potts}. This model is defined as $E(\bm{\sigma})=-\sum_{i<j}W_{ij}\delta(\sigma_i,\sigma_j)$, where $\delta(\sigma_i,\sigma_j)$ is the Kronecker function, yielding the value 1 if $\sigma_i=\sigma_j$ and 0 otherwise. 
\item  Another category of COPs includes those with higher-order interactions in the cost function, yet retain binary spin states. An example is the $p$-spin (commonly with $p>2$) Ising glass model~\cite{gardner1985spin}, characterized by the energy function $E(\bm{\sigma})=-\sum_{i_1<i_2<\ldots<i_p}W_{i_1,i_2,\ldots,i_p}\sigma_{i_1}\sigma_{i_2}\cdots\sigma_{i_p}$, which integrates interactions among $p$ distinct spins. 
This class of COPs is also known as the polynomial unconstrained binary optimization (PUBO) problem~\cite{chermoshentsev2021polynomial}. We note that considerable efforts have been undertaken to extend existing Ising solvers to high-order architectures~\cite{chermoshentsev2021polynomial,bybee2023efficient,kanao2022simulated,reifenstein2023coherent}. 
\item In more general scenarios, COPs can encompass both multi-valued states and many-body interactions, featuring a simultaneous coexistence of interactions across various orders. 
We term this class of COPs the general model, it poses more challenges for the design of extended Ising machines.
\end{enumerate}

Our proposed approach aims to address all kinds of problems discussed above (also shown in Fig.~\ref{fig:BolzmannDistribution}(a)) using the same variational framework. Within this framework, we focus on analyzing the Boltzmann distribution at a specified temperature


\begin{align}
    P_B(\bm{\sigma},\beta)=\frac{1}{Z}e^{-\beta E(\bm{\sigma})},
\end{align}
where $\beta = 1/T$ is the inverse temperature and $Z=\sum_{\bm{\sigma}}e^{-\beta E(\bm{\sigma})}$ is the partition function. It is important to emphasize that we do not impose any constraints on the specific form of 
$E(\bm{\sigma})$. Consequently, we extend the traditional Ising model formulations by permitting the spin variable to adopt 
$q$ distinct states and use $P_i(\sigma_i)$ to represent the marginal probability of the $i$-th spin taking value $\sigma_i=1,2,\cdots,q$, as illustrated in Fig.~\ref{fig:BolzmannDistribution}(b).
The ground state configuration $\bm{\sigma}^{\rm GS}$ that minimizes the energy can be achieved at the zero-temperature limit with 
\begin{align}
    \bm{\sigma}^{\rm GS}=\underset{\bm{\sigma}}{\arg\min}\ E(\bm{\sigma})=\underset{\bm{\sigma}}{\arg\max}\lim_{\beta\to\infty} P_B(\bm{\sigma},\beta).
\end{align}
As illustrated in Fig.~\ref{fig:BolzmannDistribution}(b) and (c), accessing the Boltzmann distribution at zero temperature would allow us to calculate the marginal probabilities $P_i(\sigma_i)$ and determine the configuration based on the probabilities. 
However, there are two issues to accessing the zero-temperature Boltzmann distribution. 

The first issue is that directly accessing the Boltzmann distribution at zero temperature poses significant challenges due to the rugged energy landscape, often described by the concept of replica symmetry breaking in statistical physics~\cite{mezard1984replica, mezard1987spin}. To navigate this issue and facilitate a more manageable exploration of the landscape, we employ the strategy of annealing, which deals with the Boltzmann distribution at a finite temperature. This temperature is initially set high and is gradually reduced to zero.

The second issue is how to represent the Boltzmann distribution. Exactly computing the Boltzmann distribution belongs to the computational class of \#P, so we need to approximate it efficiently. Many approaches have been proposed, including Markov-Chain Monte-Carlo~\cite{kirkpatrick1983optimization}, mean-field and message-passing algorithms~\cite{mezard2002analytic}, and neural network methods~\cite{van,hibat2021variational}. 
In this work, we use the variational mean-field distribution $P_{\mathrm MF}(\bm{\sigma})=\prod_iP_i(\sigma_i)$ to approximate the Boltzmann distribution $P_B(\bm{\sigma},\beta)$. The parameters of $\pmf$ can be determined by minimizing the Kullback-Leibler divergence $D_{\rm KL}(P\,\Vert\,P_B)=\sum_{\bm{\sigma}} P(\bm{\sigma})\ln \left(P(\bm{\sigma}) / P_B(\bm{\sigma},\beta)\right)$ ,
and this is equivalent to minimizing the variational free energy
\begin{align}\label{eq:FP}
    F_\mathrm{MF} = \sum_{\bm{\sigma}} \pmf(\bm{\sigma})E(\bm{\sigma})+\frac{1}{\beta}\sum_{\bm{\sigma}} \pmf(\bm{\sigma})\ln \pmf(\bm{\sigma}).
\end{align}
While the mean-field distribution may not boast the expressiveness of, for instance, neural network ansatzes, our findings indicate that it provides a precise representation of ground-state configurations at zero temperature via the annealing process. 
Furthermore, a significant advantage of the mean-field variational distribution is the capability for exact computation of the gradients of the mean-field free energy. This stands in stark contrast to variational distributions utilizing neural networks, where gradient computation necessitates stochastic sampling~\cite{van}.

The pictorial illustration of implementing the FEM algorithm is depicted in Fig.~\ref{fig:FEM}. Given a COP defined on graph, we associate the $N$ spin variables (each spin has $q$ states) with the variational variables represented by the $N\times q$ marginal probabilities $\{P_i(\sigma_i)\}$ for the mean-field free energy. Then we parameterize the marginal probability $P_i(\sigma_i)$ using fields $\{h_i(\sigma_i)\}$ with a softmax function $P_i(\sigma_i)=\exp[h_i(\sigma_i)]/\sum_{\sigma_i^\prime=1}^q \exp[h_i(\sigma_i^\prime)]$ for a $q$-state variable, as illustrated in Fig.~\ref{fig:FEM}(a). This parameterization can release the normalization constraints on the variational variables $\{P_i(\sigma_i)\}$ (ensuring the probabilistic interpretation of $\sum_{\sigma_i} P_i(\sigma_i)=1$ during the variational process). Moreover, our approach considers constraints on variables, such as the total number of spins with a particular value, or a global property that a configuration must satisfy.

At a high temperature, there could be just one mean-field distribution that minimizes the variational free energy. However, at a low temperature, there could be many mean-field distributions, each of which has a local free energy minimum, corresponding to a set of marginal probabilities.
Inspired by the one-step replica symmetry breaking theory of spin glasses, we use a set of marginal distributions, which we term as $m$ replicas of mean-field solutions with parameters $\{P_i^a(\sigma_i)\,|\,i=1,2,\cdots, N;\, a=1,2,\cdots,m\}$, each of which is updated to minimize the corresponding mean-field free energy and the minimization is reached through machine learning optimization techniques. This approach notably enhances both the number of parameters and the expressive capability of the mean-field ansatz. 

The parameters of the replicas of the mean-field distributions are determined by minimizing the variational free energies, the gradients can be computed using automatic differentiation. This process is very similar to computing gradients of the loss function with respect to the parameters in deep neural networks. It amounts to expanding the computational process as a computational graph and applying the backpropagation algorithm. Thanks to the standard deep learning frameworks such as PyTorch~\cite{paszke2019pytorch} and TensorFlow~\cite{abadi2016tensorflow}, it can be implemented using just several lines of code as shown in the Methods section. 
Remarkably, for different combinatorial optimization problems, we only need to specify the form of the energy expectations $\sum_{\bm{\sigma}} \pmf(\bm{\sigma}) E(\bm{\sigma})$ as a function of marginal probabilities. Beyond leveraging automatic differentiation for gradient computation, we have the option to delineate the explicit gradient formulas for the problem. Utilizing explicit gradient formulations can halve the computational time. Another merit of adopting explicit gradients lies in the possibility of further enhancing our algorithm's stability through additional gradient manipulations, beyond merely employing adaptive learning rates and momentum techniques, as seen in gradient-based optimization methods in machine learning~\cite{kingma2014adam}. 

With gradients computed, we adopt the advanced gradient-based optimization methods developed in the deep learning community for training neural networks, such as Adam~\cite{kingma2014adam}, to update the parameters. They can efficiently maintain individual adaptive learning rates for each marginal probability from the first and second moments of the gradients, and require minimal memory overhead, so is well-suited for updating marginal probabilities in our algorithm. Fig.~\ref{fig:FEM}(b) shows a schematic of the typical evolutions of internal energy $U_{\mathrm{MF}}=\sum_{\bm{\sigma}} \pmf(\bm{\sigma}) E(\bm{\sigma})$, and entropy $S_{\mathrm{MF}}=-\sum_{\bm{\sigma}} \pmf(\bm{\sigma})\ln \pmf(\bm{\sigma}) $ as a function of $\beta$. The corresponding evolutions of $\{P_i(\sigma_i)\}$ of the replicas with the annealing are depicted in Fig.~\ref{fig:FEM}(c). All mean-field probabilities of the replicas are updated parallelly.
Initially, the fields $\{h_i(\sigma_i)\}$ associated with spin $i$ are randomly initialized around zero, making $q$-state marginal distributions $\{P_i(\sigma_i)\}$ around $1/q$.
In the first stage indicated in Fig.~\ref{fig:FEM}(b), since $\beta$ is small (i.e. at a high temperature), the entropy term $S_{\mathrm{MF}}$ predominantly governs the energy landscape of $F_{\mathrm{MF}}$. The uniform distributions of $\{P_i(\sigma_i)\}$ indicate that all possible spin configurations emerge with equal importance.  Consequently, they maximize $S_{\mathrm{MF}}$ (i.e. minimize $F_{\mathrm{MF}}$ at a fixed $\beta$), and the value of $S_{\mathrm{MF}}$ remains around its maximal value in this stage. 
In the second stage, when $\beta$ increases to some critical value $\beta_c$, $U_{\mathrm{MF}}$ becomes the predominant factor and the distribution transition occurs. As a consequence, internal energy plays a more important role in minimizing $F_{\mathrm{MF}}$, leading $\{P_i(\sigma_i)\}$ to deviate from the uniform distributions and explore different mean-field solutions. 
In the third stage, when $\beta$ is sufficiently large, $U_{\mathrm{MF}}$ gradually converges to a minimum value of $F_{\mathrm{MF}}$, and $\{P_i(\sigma_i)\}$ gradually converges to approximate the zero-temperature Boltzmann distribution, where the ground states are the most probable states to occur.

After the annealing process, as shown in Fig.~\ref{fig:FEM}(d), the temperature is decreased to a very low value, and we obtain a configuration for each replica according to the marginal probabilities, as 
\begin{align}
    \widetilde{\sigma}^a_i=\underset{\sigma_i}{\arg\max}\ P_i^a(\sigma_i).
\end{align}
Then we choose the configuration with the minimum energy from all replicas
\begin{align}
\widehat{\sigma}_i=\underset{a}{\arg\min}\ E(\widetilde \sigma_i^a).
\end{align}
We emphasize that all the gradient computation and parameter updating on replicas can be processed parallelly and is very similar to the computation in deep neural networks: overall computation only involves batched matrix multiplications and element-wise non-linear functions. Thus it fully utilizes massive parallel computational devices such as GPUs.



\subsection{Applications to the Maximum-Cut problem}\par
We begin by evaluating the performance of our algorithm on Quadratic Unconstrained Binary Optimization (QUBO) problems. To illustrate, we select the MaxCut problem as a representative example. This NP-complete problem is widely applied in various fields, including machine learning and data mining~\cite{boykov2001interactive}, the design of electronic circuits~\cite{barahona1988application}, and social network analysis~\cite{facchetti2011computing}. Furthermore, it serves as a prevalent testbed for assessing the efficacy of new algorithms aimed at solving QUBO problems~\cite{goto2019combinatorial,goto2021high,bohm2021order}. 
The optimization task in the MaxCut problem is to determine an optimal partition of nodes into $q=2$ groups for an undirected weighted graph, in such a way that the sum of weights on the edges connecting two nodes belonging to different partitions (i.e. the cut size $C$) is maximized. Formally, we define the energy function as
\begin{align}
\label{eq:maxcut1}
    E(\bm{\sigma}) = -\sum_{(i,j)\,\in\,\mathcal{E}}W_{ij}[1-\delta(\sigma_i,\sigma_j)],
\end{align}
where $\mathcal{E}$ is the edge set of the graph, $W_{ij}$ is the weight of edge $(i,j)$, and $\delta(\sigma_i,\sigma_j)$ stands for the delta function, which takes value 1 if $\sigma_i=\sigma_j$ and 0 otherwise.
Then the variational mean-field free energy for the MaxCut problem can be written out (see Methods section) and the gradients of $F_{\mathrm{MF}}$ to the variational parameters can be computed via automatic differentiation. In general, writing out the explicit formula for the gradients is not necessary, as the gradients computed using automatic differentiation are numerically equal to the explicit formulations. However, for the purpose of benchmarking, using the explicit formula will result in lower computation time. Moreover, in practice, we can further apply the normalization and clip of gradients to enhance the robustness of the optimization process, we refer to the Supplementary Materials for details.

We first benchmark FEM by solving a 2000-spin MaxCut problem named $K_{2000}$~\cite{k2000} with all-to-all connectivity, which has been intensively used in evaluating MaxCut solvers~\cite{inagaki2016coherent,goto2019combinatorial,goto2021high}.  We compare the results obtained by FEM with discrete-SBM (dSBM)~\cite{goto2021high}, which can be identified as the state-of-the-art solver for the MaxCut problem. 
Fig.~\ref{fig:maxcut}(a) shows the cut values we obtained for the $K_{2000}$ problem with different total annealing steps $N_{\rm step}$ used to increase $\beta$ from $\beta_{\rm min}$ to $\beta_{\rm max}$. Similar to the role in FEM, the hyperparameter $N_{\rm step}$ introduced in dSBM controls the total number of annealing steps for the bifurcation parameter. To investigate the distribution of energy in all replicas, in the figure, we plot the minimum, average, and maximum cut value as a function of $N_{\rm step}$, with $R=1000$ mean-field replicas for FEM. The results of FEM are compared with the dSBM algorithm, for which we also ran for $1000$ trials from random initial conditions. From the results, we can see that the best value of FEM achieves the best-known results for this problem in less than $1000$ annealing steps, and all the maximum, average, and minimum cut sizes are better than dSBM.

We also evaluate our algorithm using the standard benchmark for the MaxCut problem, the G-set~\cite{gset,inagaki2016coherent,goto2021high}. The G-set benchmark contains various graphs including random, regular, and scale-free graphs based on the distribution of node degrees as shown in Fig.~\ref{fig:maxcut}(b). Each problem contains a best-known solution which is regarded as the ground truth for evaluating algorithms. A commonly used statistic for quantitatively assessing both the accuracy and the computational speed in the MaxCut problem is the time-to-solution (TTS)~\cite{goto2021high,bohm2021order}, which measures the average time to find a good solution by running the algorithm for many times (trials) from different initial states. TTS (or ${\rm TTS}_{99}$) is formulated as $T_{\rm com}\cdot \log(1-0.99)/\log (1-P_S)$, where $T_{\rm com}$ represents the computation time per trial, and $P_S$ denotes the success probability of finding the optimal cut value in all tested trials. When $P_S\ge0.99$, the TTS is defined simply as $T_{\rm com}$.
The value of $P_S$ is typically estimated from experimental results comprising numerous trials. 
\begin{table*}[htb]
    \centering
    \resizebox{\textwidth}{!}{%
    \begin{tabular}{@{\hspace{0.5em}}c@{\hspace{0.5em}}|@{\hspace{0.5em}}c@{\hspace{0.5em}}|@{\hspace{0.5em}}c@{\hspace{0.5em}}|@{\hspace{0.5em}}c@{\hspace{0.5em}}|@{\hspace{0.5em}}c@{\hspace{0.5em}}|@{\hspace{0.5em}}c@{\hspace{0.5em}}"@{\hspace{0.5em}}c@{\hspace{0.5em}}|@{\hspace{0.5em}}c@{\hspace{0.5em}}|@{\hspace{0.5em}}c@{\hspace{0.5em}}|@{\hspace{0.5em}}c@{\hspace{0.5em}}|@{\hspace{0.5em}}c@{\hspace{0.5em}}|@{\hspace{0.5em}}c@{\hspace{0.5em}}}
       \Xcline{1-12}{0.8pt}
         \multirow{2}*{Instance}&\multirow{2}*{$\ q\ $}&\multirow{2}*{\makecell[c]{Best\\known}} & \multirow{2}*{METIS} &\multirow{2}*{KaFFPaE}& \multirow{2}*{FEM}&\multirow{2}*{Instance}& \multirow{2}*{$\ q\ $}&\multirow{2}*{\makecell[c]{Best\\known}} & \multirow{2}*{METIS} &\multirow{2}*{KaFFPaE}& \multirow{2}*{FEM}\\
         ~&~&~&~&~&~&~&~&~&~&~&~\\
        \Xcline{1-12}{0.5pt}
    
         \multirow{5}*{add20}&2&596(0)&722(0)&597(0)&\textbf{596(0)}&\multirow{5}*{data}&2&189(0)&211(0)&\textbf{189(0)}&\textbf{189(0)}\\
      
        ~&4&1151(0)&1257(0)&1158(0)&\textbf{1152(0)}&~&4&382(0)&429(0)&\textbf{382(0)}&\textbf{382(0)}\\
      
        ~&8&1678(0)&1819(0.007)&1693(0)&\textbf{1690(0)}&~&8&668(0)&737(0)&\textbf{668(0)}&669(0)\\
 
        ~&16&2040(0)&2442(0)&\textbf{2054(0)}&2057(0)&~&16&1127(0)&1237(0)&1138(0)&\textbf{1129(0)}\\

        ~&32&2356(0)&2669(0.04)&2393(0)&\textbf{2383(0)}&~&32&1799(0)&2023(0)&1825(0)&\textbf{1815(0)}\\

        
        \Xcline{1-12}{0.5pt}
        \multirow{5}*{3elt}&2&90(0)&\textbf{90(0)}&\textbf{90(0)}&\textbf{90(0)}&\multirow{5}*{bcsstk33}&2&10171(0)&10205(0)&\textbf{10171(0)}&\textbf{10171(0)}\\
        ~&4&201(0)&208(0)&\textbf{201(0)}&\textbf{201(0)}&~&4&21717(0)&22259(0)&\textbf{21718(0)}&\textbf{21718(0)}\\
      
        ~&8&345(0)&380(0)&\textbf{345(0)}&\textbf{345(0)}&~&8&34437(0)&36732(0.001)&\textbf{34437(0)}&34440(0)\\
  
        ~&16&573(0)&636(0.004)&\textbf{573(0)}&\textbf{573(0)}&~&16&54680(0)&58510(0)&54777(0)&\textbf{54697(0)}\\
 
        ~&32&960(0)&1066(0)&966(0)&\textbf{963(0)}&~&32&77410(0)&83090(0.004)&77782(0)&\textbf{77504(0)}\\
        \Xcline{1-12}{0.5pt}
        
    \end{tabular}
}
    \caption{\textbf{Benchmarking results for the $q$-way balanced minimum cut problem, using real-world graph instances from Chris Walshaw’s archive~\cite{walshawgraph}.} 
The graph partitioning was tested with the number of groups $q$ set to 2, 4, 8, 16, and 32 across all solvers. The table showcases the minimum cut values obtained by three different solvers. The numbers in parentheses represent the imbalance
$\epsilon$ (with $\epsilon = 0$ being ideal) satisfying the condition $|\Pi_n|\le(1+\epsilon)\lceil N/q \rceil$ for $n=1,2,\ldots,q$, where $N$ denotes the number of nodes and $|\Pi_n|$ indicates the size of each group. For each graph, the lowest cut values are emphasized in bold.}
    \label{tab:bmincut_more}
\end{table*}

In Fig.~\ref{fig:maxcut}(c) we present the TTS results obtained by FEM for $54$ problem instances in G-set (G1 to G54, with the number of nodes ranging from 800 to 2000) and compare to the reported data for dSBM using GPUs.
We can see that FEM surpasses dSBM in TTS for 33 out of the 54 instances, and notably for all G-set instances of scale-free graph, FEM achieves better performance than dSBM. The primary reason is that we employed the advanced normalization techniques for the optimization, please refer to the Supplementary Materials for more details.  
Furthermore, we notice that FEM outperforms the state-of-the-art neural network-based method in combinatorial optimization. For instance, the physics-inspired graph neural network (PI-GNN) approach \cite{schuetz2022combinatorial} has been shown to outperform other neural-network-based methods on the G-set dataset. From the data in \cite{schuetz2022combinatorial} we see that results of PI-GNN on the G-Set instances (with estimated runtimes in the order of tens of seconds) still give significant discrepancies to the best-known results of the G-set instances e.g. for G14, G15, G22 etc, while our method FEM achieves the best-known results for these instances only using several milliseconds.

\subsection{Applications to the $q$-way balanced minimum cut problem}\par

Next, we choose the $q$-way balanced MinCut (bMinCut) problem~\cite{ushijima2017graph} as the second benchmarking type to evaluate the performance of FEM on directly addressing multi-valued problems featuring the Potts model. The $q$-way bMinCut problem asks to group nodes in a graph into $q$ groups with a minimum cut size and balanced group sizes. The requirement to balance group sizes imposes a global constraint on the configurations, rendering the problem more complex than unconstrained problems such as the MaxCut problem. 
Here we formulate the energy function with a soft constraint term
\begin{equation}
\label{eq:bmincut}
    E(\bm{\sigma}) = \sum_{(i,j)\,\in\,\mathcal{E}}W_{ij}[1-\delta(\sigma_i,\sigma_j)] + \lambda\sum_i\sum_{j\neq i} \delta(\sigma_i, \sigma_j),
\end{equation}
where $\lambda$ is the parameter of soft constraints which controls the degree of imbalance.
Based on the energy formulation, the expression of $F_{\rm MF}$ can be explicitly formulated (see Methods section), and its gradients can be calculated through automatic differentiation or derived analytically. In practice, we also used gradient normalization and gradient clip to enhance the robustness of the optimization, we refer to Supplementary Materials for detailed discussions.
It's worth noting that the bMinCut problem bears a significant resemblance to the community detection problem~\cite{newman2006modularity}. In the latter, the imbalance constraint is often substituted with the constraints derived from a random configuration model~\cite{newman2006modularity} or a generative model~\cite{holland1983stochastic, karrer2011stochastic}, to avoid the trivial solution that puts all the nodes into a single group. Thus FEM can be easily adapted to the community detection problem.



\begin{figure*}[htbp]
\centering
\includegraphics[width=1\linewidth]{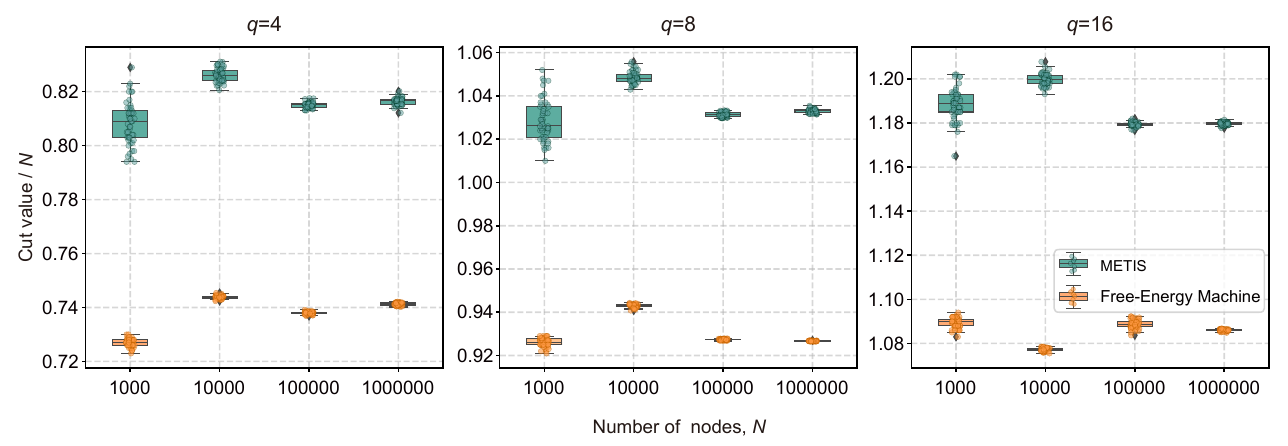}
\caption{\textbf{The scalability test of FEM in the $q$-way bMinCut problems on the Erd\"os-R\'enyi random graphs.} The generated random graphs contain a number of nodes ranging from 1,000 to 1,000,000, each with an average degree of 5. We benchmark the scalability of FEM with METIS over these generated random graphs. The number of partitions we evaluate are $q=2, 4, 8$. Each box plot contains 50 points, representing the cut value per node, obtained from 50 runs of the algorithms. The number of replicas of FEM is set to $R=50$ (in each run) throughout the experiments. } 
\label{fig:largeN}
\end{figure*}

\begin{figure*}[htbp]
\centering
\includegraphics[width=0.95\linewidth]{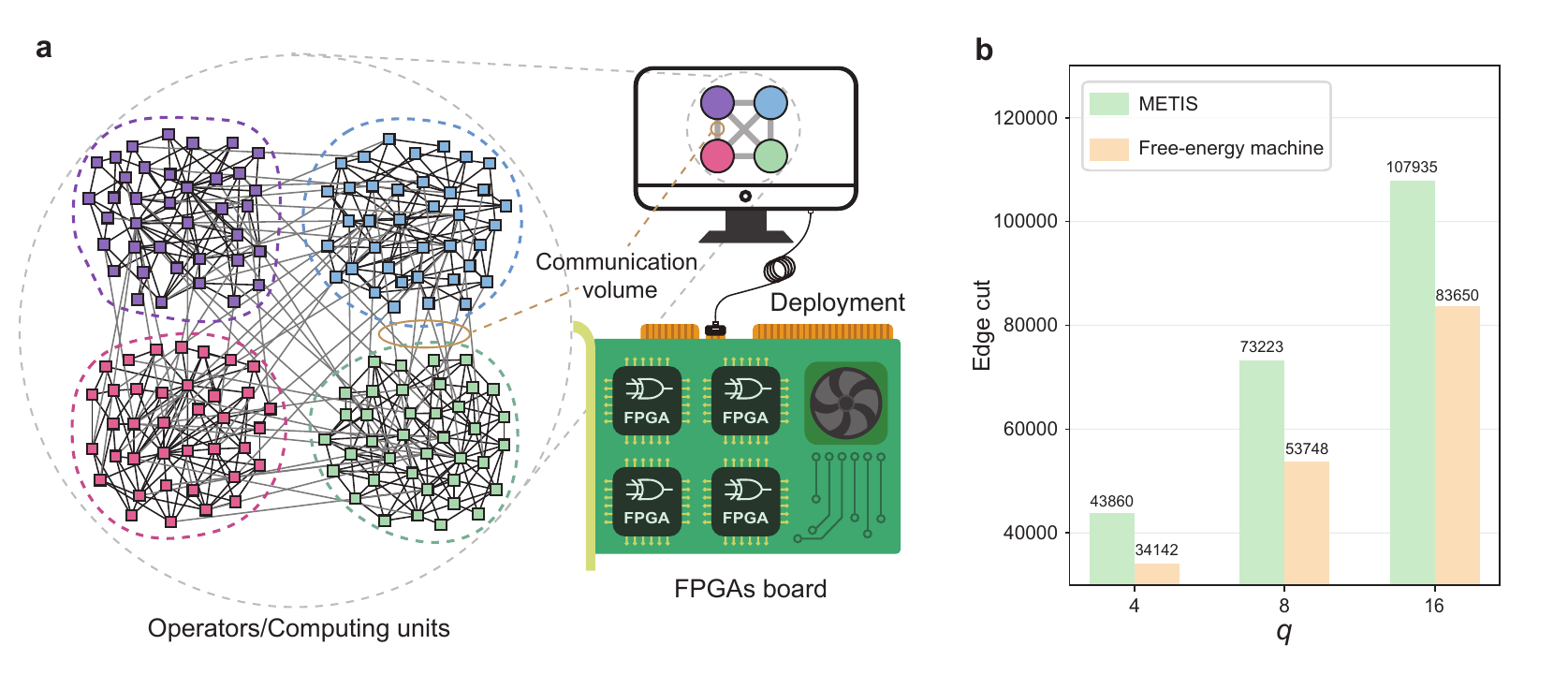}
\caption{{\textbf{The application of FEM in large-scale FPGA-chip verification tasks.} \textbf{(a).} In the chip verification task, a vast number of operators or computing units with logical interconnections need to be uniformly deployed onto a hardware platform consisting of several FPGAs. The operators are partitioned into several groups corresponding to different FPGAs, and the communication volume between these groups of operators should be minimized. This can be modeled as a balanced minimum cut problem. \textbf{(b).} The results of FEM, along with a comparison to METIS, are presented for a large-scale real-world dataset consisting of 1,495,802 operators and 3,380,910 logical operator connections deployed onto ($q$ = 4, 8, 16)  FPGAs (see Supplementary Materials for details).}} 
\label{fig:FPGA}
\end{figure*}

To evaluate the performance of FEM in solving the $q$-way bMinCut problem, we conduct numerical experiments using four large real-world graphs from Chris Walshaw's archive~\cite{walshawgraph}. These include \emph{add20} with 2395 nodes and 7462 edges, \emph{data} with 2851 nodes and 15093 edges,  \emph{3elt} which comprises 4,720 nodes and 13,722 edges, and \emph{bcsstk33} with 8738 nodes and 291583 edges.  
The graphs have been widely used in benchmarking $q$-way bMinCut solvers, e.g. used by the D-wave for benchmarking their quantum annealing hardware~\cite{ushijima2017graph}. However, their work only presents the results of $q=2$ partitioning, owing to the constraints of the quantum hardware. 
Here, we focus on the perfectly balanced problem which asks for group sizes the same. We evaluate the performance of FEM by partitioning the graphs into $q=2, 4, 8, 16, 32$ groups.
For comparison, we utilized two state-of-the-art solvers tailored to the $q$-way bMinCut problem: METIS~\cite{karypis1998fast} and KAHIP (alongside its variant KaFFPaE, specifically engineered for balanced partitioning)~\cite{sanders2013think}, the latter being the winner of the 10th DIMACS challenge.
The benchmarking results are shown in Tab.~\ref{tab:bmincut_more}, where we can observe that the results obtained by FEM consistently and considerably outperform METIS in all the problems and for all the number of groups $q$. 
Moreover, in some instances, METIS failed to find a perfectly balanced solution while the results found by FEM in all cases are perfectly balanced. 
We observe that FEM performs comparably to KaFFPaE for small group sizes $q$, and significantly outperforms KaFFPaE with a large $q$ value. 
We have also evaluated the performance of FEM on extensive random graphs comprising up to one million nodes. The outcomes are depicted in Fig.~\ref{fig:largeN}.  As observed from the figure, FEM achieves significantly lower cut values than METIS across the same collection of graphs when partitioned into $q=4, 8$, and $16$ groups. Notably, this performance disparity is maintained as the number of nodes increases to one million.
The comparisons demonstrate FEM's exceptional scalability in solving the large-scale $q$-way bMinCut problems.

Since the bMinCut modeling finds many real-world applications in parallel computing and distributed systems, data clustering, and bioinformatics~\cite{acer2021sphynx,chuzhoy2020deterministic}, we then apply FEM to address a challenging real-world problem of chip verification~\cite{lam2005hardware}. To identify and correct design defects in the chip before manufacturing, operators or computing units need to be deployed on a hardware platform consisting of several processors (e.g. FPGAs) for logic and function verification. Due to the limited capacity of a single FPGA and the restricted communication bandwidth among FPGAs, a large number of operators need to be uniformly distributed across the available FPGAs, while minimizing the communication volume among operators on different FPGAs. The schematic illustration is shown in Fig.~\ref{fig:FPGA}(a), This scenario resembles load balancing in parallel computing and minimizing the edge cut while maintaining balanced partitions and can be modeled as a balanced minimum cut problem~\cite{patil2021k}. 
In this work, we address a large-scale real-world chip verification task consisting of 1495802 operators (viewed as nodes) and 3380910 logical operator connections (viewed as edges) onto $q=4, 8, 16$ FPGAs. We apply FEM to solve this task. Since the dataset contains many locally connected structures among operators, we first conduct coarsening the entire graph before partitioning on it. Unlike the matching method used in the coarsening phase in METIS~\cite{karypis1998fast}, we apply the Louvain algorithm~\cite{blondel2008fast} to identify community structures. Nodes within the same community are coarsened together. The results are shown in Fig.~\ref{fig:FPGA}(b), along with comparative results provided by METIS (see Supplementary Materials for more details. We did not include the results of KaFFPaE, as its open-source implementation~\cite{kahipgithub} runs very slowly and exceeds the acceptable time limits on large-scale graphs). From the figure, we can observe that 
the size of the edge cut given by FEM is 22.2\%, 26.6\%, and 22.5\% smaller than METIS, for 4,8, and 16 FPGAs, respectively, significantly reduces the amount of communication among FPGAs and shortening the chip verification time.

\begin{figure*}[htbp]
    \centering
    \includegraphics[width=1\linewidth]{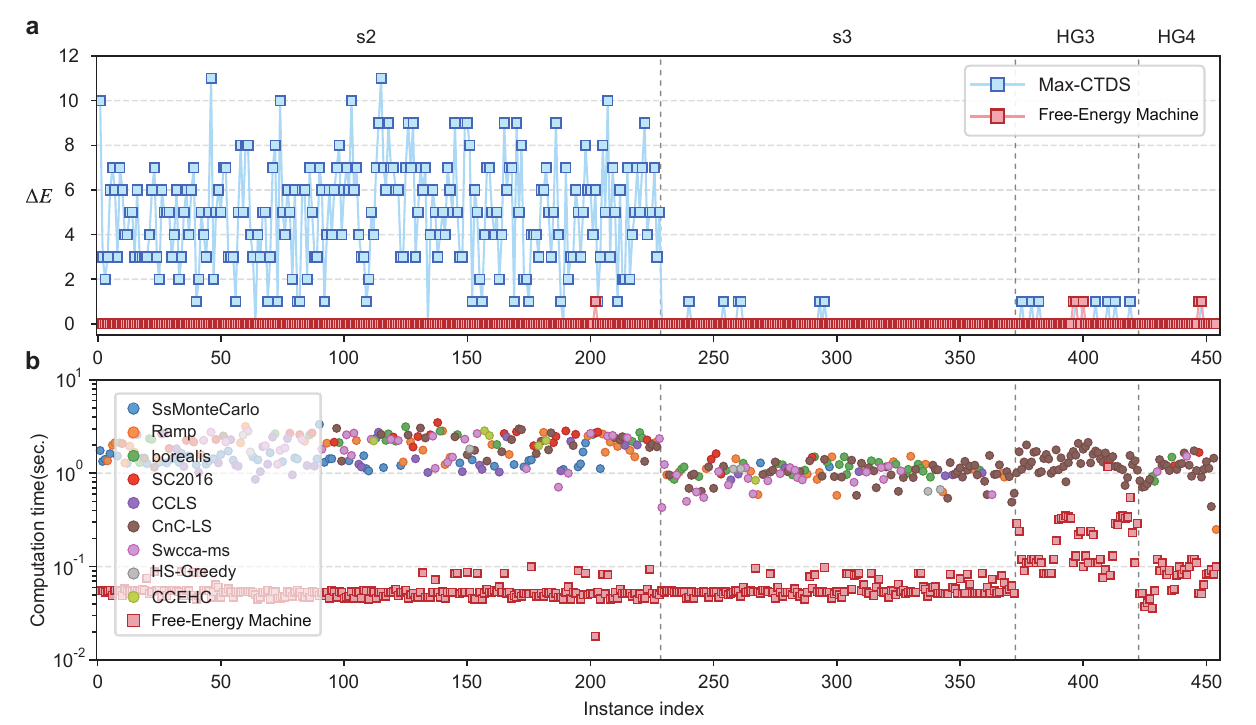}
    \caption{\textbf{Benchmarking results on the MaxSAT 2016 competition problems.}  \textbf{(a)}. The results for the energy differences, $\Delta E =E_{\rm min}- E_{\rm bkr}$, which represent the gap between the minimal energy values found by the solvers ($E_{\rm min}$) and the best-known results documented in the literature ($E_{\rm bkr}$) with $E$ indicating the number of unsatisfied clauses. These findings cover all 454 problem instances across four competition categories: ``s2'', ``s3'', ``HG3'', and ``HG4''. For further discussion on these results, please see the main text. The Max-CTDS algorithm data were obtained from Ref.~\cite{molnar2018continuous}. \textbf{(b)}. 
    The computation time (measured by the running time of all replicas) of the Free-Energy Machine (FEM) to achieve these outcomes against the leading incomplete solvers from the 2016 competition across different instances. The performance data for these incomplete solvers were sourced from the 2016 MaxSAT competition documentation~\cite{maxksat2016}. For detailed information on the experimental setup for FEM, refer to the main text and Supplementary Materials.}
    \label{fig:maxksat2}
\end{figure*}

\subsection{Application to the Max $k$-SAT problem}
Lastly, we evaluate FEM for addressing COPs
with the higher-order spin interactions on the constraint Boolean satisfiability (SAT) problem. In this problem, $M$ logical clauses, denoted as $C_1, C_2, \ldots, C_M$, are applied to $N$ boolean variables. Each clause is a disjunction of literals, namely $C_m=l_1\lor l_2\lor \ldots \lor l_{k_m}$($k_m$ is the number of literals in the  clause $C_m$). A literal can be a Boolean variable $\sigma_i$ or its negation $\neg \sigma_i$. The clauses 
are collectively expressed in the conjunctive normal form (CNF). For example, the CNF formula $C_1\land C_2 \land C_3 =(\sigma_1\lor \neg \sigma_2 \lor \neg \sigma_2 ) \land (\neg \sigma_1 \lor \sigma_4) \land (\sigma_2\lor \neg \sigma_3\lor \neg \sigma_4)$ is composed of 4 Boolean variables and 3 clauses. Note that, a clause is satisfied if at least one of its literals is true, and is unsatisfied if no literal is true. We can see that the SAT problem is a typical many-body interaction problem with higher-order spin interactions.
The decision version of the SAT problem asks to determine whether there exists an assignment of Boolean variables to satisfy all clauses simultaneously (i.e. the CNF formula is true). The optimization version of the SAT problem is the maximum SAT 
(Max-SAT) problem, which asks to find an assignment of variables that maximizes the number of clauses that are satisfied.
When each clause comprises exactly $k$ literals (i.e. $k_m=k$), the problem is identified as the $k$-SAT problem, one of the earliest recognized NP-complete problems (when $k\ge 3$) ~\cite{10.1145/800157.805047, cook2000p}. These problems are pivotal in the field of computational complexity theory. 
We benchmark FEM on the Max $k$-SAT problem, which is NP-hard for any $k \geq 2$.
In our framework, the energy function for the Max $k$-SAT problem is formulated as the number of unsatisfied clauses, as
\begin{align}
\label{eq:maxksat_E}
    E(\bm{\sigma}) = \sum_{m=1}^M\prod_{i\in\partial m}[1-\delta (W_{mi},\sigma_i)] \; ,
\end{align}
where $\sigma_i=\{0, 1\}$ is the Boolean variable, $\partial m$ denotes the set of Boolean variables that appears in clause $C_m$, $W_{mi}=0$ if the literal regarding variable $i$ is negated in clause $C_m$ and $W_{mi}=1$ when not negated. Note that, in the case of Boolean SAT, $W_{mi}\in \{0,1\}$ corresponds to the two states of spin variables. The energy function can be generalized to any Max-SAT problem, where clauses may vary in the number of literals, and to cases of non-Boolean SAT problem where $W_{mi}$ has $q>2$ states. The expression of $F_{\rm MF}$ can be also found in Methods section, and the form of its explicit gradients please refer to Supplementary Materials.


We access the performance of FEM using the dataset in MaxSAT 2016 competition~\cite{maxksat2016}. The competition problems encompass four distinct categories: ``s2'', ``s3'' by Abrame-Habet, and ``HG3'', ``HG4'' with high-girth sets. The ``s2'' category consists of Max 2-SAT problems with $N \in [120, 200]$ and $M \in [1200, 2600]$. The ``s3'' category includes Max 3-SAT problems with $N \in [70, 110]$ and $M \in [700, 1500]$. For the ``HG3'' category, the Max 3-SAT problems feature $N \in [250, 300]$ and $M \in [1000, 1200]$. Lastly, the ``HG4'' category contains Max 4-SAT problems with $N \in [100, 150]$ and $M \in [900, 1350]$. Fig.~\ref{fig:maxksat2} shows the benchmarking results for all 454 competition instances in the four categories (see Supplementary Materials for the experimental details).  

In Fig.~\ref{fig:maxksat2}(a), we illustrate the quality of solutions found by FEM, evaluated using the energy difference $\Delta E$, between FEM the best-known results for all problem instances. To provide a comprehensive comparison, we also present the documented results for the competition problems achieved by a state-of-the-art solver, continuous-time dynamical heuristic (Max-CTDS), as reported in Ref.~\cite{molnar2018continuous}. 
The results in Fig.~\ref{fig:maxksat2}(a) show that FEM found the optimal solution in 448 out of 454 problem instances. For the $6$ instances that FEM did not achieve the optimal solution, it found a solution with an energy gap $1$ to the best-known solution. Also, we can see that FEM outperforms Max-CTDS in all instances.

In Fig.~\ref{fig:maxksat2}(b), we list the computational time of FEM using GPU in solving each instance of SAT competition 2016 problems and compare them with the computation time of the specific-purpose incomplete MaxSAT solvers (using CPU) in the 2016 competition~\cite{maxksat2016}. 
For each instance, we only chart the minimum computational time of the incomplete solver needed to reach the best-known results, as documented by all incomplete MaxSAT solvers~\cite{maxksat2016}. Note that the fastest incomplete solver can differ across various instances.
The data presented in the figure clearly demonstrates that FEM outperforms the quickest incomplete MaxSAT solvers from the competition, both significantly and consistently. On average, FEM achieves a computational time of 0.074 seconds across all instances, with a variation (standard deviation) of 0.077 seconds. The computational time ranges from as short as 0.018 seconds for the ``s2v200c1400-2.cnf'' to as long as 1.17 seconds for the ``HG-3SAT-V250-C1000-14.cnf'' instance.  
A key factor contributing to the rapid computation time of FEM is its ability to leverage the extensive parallel processing capabilities of GPU, which can accelerate computations by approximately tenfold compared to CPU processing. Nevertheless, it's crucial to highlight that even when performing on a CPU, FEM significantly outpaces most of the competitors in the SAT competition. For example, on average, Max-CTDS demands an average time of 4.35 hours to approximate an optimal assignment across all instances, as reported in~\cite{molnar2018continuous}. In stark contrast, FEM, when running on CPU, completes the same task in just a few seconds on average. Our benchmarking results demonstrate that FEM surpasses contemporary leading solvers in terms of accuracy and computational speed when addressing the problems presented in the MaxSAT 2016 competition. 
\section{Discussion\label{sec:con}}
We have presented a general and high-performance approach for solving COPs using a unified framework inspired by statistical physics and machine learning.
The proposed method, FEM, integrates three critical components to achieve its success. First, FEM employs the variational mean-field free energy framework from statistical physics. The framework facilitates the natural encoding of diverse COPs, including those with multi-valued states and higher-order interactions. This attribute renders FEM an exceptionally versatile approach. 
Second, inspired by replica symmetry breaking theory, FEM maintains a large number of replicas of mean-field free energies, exploring the mean-field spaces to efficiently find an optimal solution. Third, the mean-field free energies are computed and minimized using machine-learning techniques including automatic differentiation, gradient normalization, and optimization. This offers a general framework for different kinds of COPs, enables massive parallelization using modern GPUs, and fast computations.

We have executed comprehensive benchmark tests on a variety of optimization challenges, each exhibiting unique features. These include the MaxCut problem, characterized by a two-state and two-body interaction without constraints; the bMinCut, defined by a $q$-state and two-body interaction with global constraints; and the Max $k$-SAT problem, which involves a two-state many-body interactions. The outcomes of our benchmarks clearly show that FEM markedly surpasses contemporary algorithms tailored specifically for each problem, demonstrating its superior performance across these diverse optimization scenarios.
Beyond the benchmarking problems showcased in this study, we also extend our modelings to encompass a broader spectrum of combinatorial optimization issues, to which FEM can be directly applied. For further details, please consult the Supplementary Materials.

In this study, our exploration was confined to the most fundamental mean-field theory within statistical physics. However, more sophisticated mean-field theories exist, such as the Thouless-Anderson-Palmer (TAP) equations associated with TAP free energy, and belief propagation, which connects to the Bethe free energy. These advanced theories have the potential to be integrated into the FEM framework, offering capabilities that could surpass those of the basic mean-field approaches. We put this into future work.

\section{Methods}
\subsection{The variational mean-field free energy formulations}

As outlined in the opening of the Results section, FEM addresses COPs by minimizing the variational mean-field free energy through a process of annealing from high to low temperatures. To tackle a specific COP, we commence by constructing the variational mean-field free energy formulation for the problem at hand. Here, we establish the variational mean-field free energy formulations for the MaxCut, the bMinCut, and the Max $k$-SAT problems that are benchmarked in this study. The derivation details can be found in Supplementary Materials.

Starting with the MaxCut problem, the variational free energy reads
\begin{align}
    F^{\rm MaxCut}_{\rm MF}(\{P_i(\sigma_i)\},\beta) &= -\sum_{(i,j)\,\in\,\mathcal {E}}\sum_{\sigma_i}W_{ij}P_i(\sigma_i)[1-P_j(\sigma_i)] \nonumber \\
    &+\frac{1}{\beta}\sum_i\sum_{\sigma_i}P_i(\sigma_i)\ln P_i(\sigma_i).
\end{align}
For the bMinCut problem, the variational free energy is written as
\begin{align}
\label{eq:loss:bmincut}
    F^{\rm bMinCut}_{\rm MF}(\{P_i(\sigma_i)\},\beta) &= \sum_{(i,j)\, \in\,\mathcal E}\sum_{\sigma_i}W_{ij}P_i(\sigma_i)[1-P_j(\sigma_i)]\nonumber\\ 
    &+ \lambda\sum_{i,j}\sum_{\sigma_i}[P_i(\sigma_i)P_j(\sigma_i)-P^2_i(\sigma_i)]\nonumber\\
    &+\frac{1}{\beta}\sum_i\sum_{\sigma_i}P_i(\sigma_i)\ln P_i(\sigma_i),
\end{align}
and for the Max $k$-SAT problem, as
\begin{align}
\label{eq:maxksat_F}
    F^{{\rm MaxSAT}}_{\rm MF}(\{P_i(\sigma_i)\},\beta) &= \sum_{m=1}^M\prod_{i\in\partial m}\left[1-P_i(W_{mi})\right]\nonumber\\ 
   &+\frac{1}{\beta}\sum_i\sum_{\sigma_i}P_i(\sigma_i)\ln P_i(\sigma_i). 
\end{align}
\subsection{Implementation of FEM with the automatic differentiation by Pytorch}
The gradients to the marginal probabilities can be computed via automatic differentiation method, or the explicit formulations (see Supplementary Materials).
Then we employ the gradient-based optimization methods, such as stochastic gradient descent (SGD), RMSprop~\cite{tieleman2012lecture}, and Adam~\cite{kingma2014adam} to update the marginal distributions and the fields, then decrease the temperature. The Python code below demonstrates a clear and direct approach to solving both the MaxCut problem and the bMinCut problem, which are as the benchmarking problems in the work. Interestingly, despite the significant differences between these two problems in terms of the number of variable states, the presence of constraints, and the nature of the objective function, the implementations for each problem differ by only a single line of code. This highlights the adaptability and efficiency of the approach in handling distinct optimization challenges. Please refer to Supplementary Materials for the codes with detailed explanations.

\label{examplecode}\label{code}
\begin{lstlisting}[language=Python]
import torch

def cut(W,p):
    return ((W @ p) * (1-p)).sum((1, 2))
    
def S(p):
    return -(p*p.log()).sum(2).sum(1)
    
def argmax_cut(W,p):
    s = torch.nn.functional.one_hot(
    p.argmax(dim=2), num_classes=p.shape[2])
    return config, cut(W, s) / 2

def balance(p):
    return (p.sum(1)**2).sum(1)-(p**2).sum(2).sum(1)
    
def solve(problem,W,batch,q,panelty,beta_range):
    n = W.shape[0]
    h = torch.rand(batch, n, q)
    optimizer = torch.optim.Adam([h], lr=0.01)
    for beta in beta_range:
        p = torch.softmax(h, dim=2)
        if problem == 'maxcut':
            F = -cut(W,p) - S(p)/beta
        if problem == 'bmincut':
            F = cut(W,p) 
                  +panelty*balance(p)-S(p)/beta
        optimizer.zero_grad()
        F.backward(gradient=torch.ones_like(F))
        optimizer.step()
    return argmax_cut(W, p)
\end{lstlisting}
As highlighted in the Results section, besides using automatic differentiation for gradient computation, we can specify explicit gradient formulas for each problem. Our numerical experiments show that employing explicit gradients can reduce computational time by half and it offers the potential to improve our algorithm's stability.
For instance, in the maximum cut problem, the variance in node degrees within the graph can lead to significant fluctuations in the gradients for each marginal, potentially destabilizing the optimization process. To counteract this, we substitute the marginal distributions with one-hot vectors and normalize the gradient magnitude for each spin, ensuring the gradients remain stable and neither explode nor vanish. For technical specifics, we direct readers to the Supplementary Materials.

\subsection{Different annealing schedules for the inverse temperature}
Regarding the annealing process, this study employs two monotonic functions to structure the annealing schedule of $\beta$. The first function is named as the exponential scheduling, which is utilized for the exponential decrease of temperature $T$. This is defined as follows:
\[\beta(t_s)= \exp\left(\frac{\ln{\beta_{\rm max}}-\ln{\beta_{\rm min}}}{N_{\rm step}-1}t_s + \ln{\beta_{\rm min}}\right),\] 
where $t_s \in \{0, 1, 2, \ldots, N_{\rm step}-1\}$ represents the annealing step within a total of $N_{\rm step}$ steps, ensuring that $\beta(0) = \beta_{\rm min}$ and $\beta(N_{\rm step}-1) = \beta_{\rm max}$. 
The second function is named as the inverse-proportional scheduling, as \[\beta(t_s)=\left(\frac{\beta_{\rm max}^{-1}-\beta_{\rm min}^{-1}}{N_{\rm step}-1} t_s+\beta_{\rm min}^{-1}\right)^{-1},\] 
which is equivalent to linear cooling in terms of temperature, with \[T(t_s)=1/\beta(t_s)=\frac{T_{\rm min}-T_{\rm max}}{N_{\rm step}-1} t_s+T_{\rm max},\] 
where $T_{\rm min}=1/\beta_{\rm max}$ and $T_{\rm max}=1/\beta_{\rm min}$.

\subsection{Hyperparameter tunning of different optimizers used for the optimization}
In this study, we employ SGD, RMSprop, and Adam as the principal optimization algorithms for our tasks. We utilize the Python implementations of these optimizers as provided by PyTorch~\cite{optimPytorch}. Within PyTorch, the adjustable hyperparameters for these optimizers vary to some extent. We have carefully tuned these hyperparameters to enhance performance while maintaining the default settings for other parameters provided by PyTorch. Specifically, for SGD, we have optimized the learning rate, weight decay, momentum, and dampening. For RMSprop, we have adjusted the learning rate, weight decay, momentum, and the smoothing constant alpha. Lastly, for Adam, the learning rate, weight decay, and the exponential decay rates for the first and second moment estimates, $\beta_1$ and $\beta_2$, have been the primary focus of our tuning efforts.

In our numerical experiments, we observed that the mean value of the energy function for COP largely depends on the hyperparameters of the optimizer used, showing a great insensitivity to the number of replicas. Consequently, hyperparameter tuning can initially be conducted with a small number of replicas, followed by incrementally increasing the number of replicas to identify the better energy values. For details on how increasing the number of replicas impacts the energy function values, please refer to the Supplementary Materials.
\subsection{Relationship to the existing mean-field annealing approaches}
It's noteworthy that the exploration of mean-field theory coupled with an annealing scheme for COPs began in the late 20th century, as indicated in ~\cite{bilbro1988optimization}. This approach has also been instrumental in deciphering the efficacy of recently introduced algorithms inspired by quantum dynamics, as discussed in~\cite{king2018emulating}. Traditional mean-field annealing algorithms, those addressing the Ising problem, revolve around the iterative application of mean-field equations (for reproducing the mean-field equations for the Ising problem from the FEM formalism, please refer to Supplementary Materials):
\[m_i = \tanh(\beta\sum_{j}W_{ij}m_j),\] 
which determines the state of a spin based on the average value of neighboring spins. These algorithms incorporate a damping technique to enhance the convergence of the iterative equations, represented as:
\[m_i^{t+1} = \alpha \, m_i^{t} + (1-\alpha)\tanh(\beta\sum_{j}^NW_{ij}m_j^{t}),\] 
where $0<\alpha<1$. In contrast, our algorithm adopts a more statistical-physics-grounded approach, focusing on the direct minimization of replicas of mean-field free energies through contemporary machine-learning methodologies. Furthermore, whereas existing iterative mean-field annealing algorithms are tailored specifically to two-state Ising problems, our algorithm boasts broader applicability, seamlessly addressing a wide array of COPs characterized by multiple states and multi-body interactions. 

\section*{Data availability}
The datasets utilized in this study for benchmarking the MaxCut problem, namely the $K_{2000}$ and G-set, were obtained from Refs.~\cite{k2000} and~\cite{gset}, respectively. For the bMinCut problem benchmarks, graph instances were sourced from Ref.~\cite{walshawgraph}. Additionally, the dataset employed for the MaxSAT benchmarks was acquired from the MaxSAT 2016 competition, as documented in Ref.~\cite{maxksat2016}.

\section*{Code availability}
The source code for this paper is publicly available at \href{HTTPs://...}{\href{https://github.com/Fanerst/FEM}{https://github.com/Fanerst/FEM}}.

\section*{Acknowledgements}
This work is supported by Project  12047503, 12325501, and 12247104 of the National Natural Science Foundation of China and project ZDRW-XX-2022-3-02 of the Chinese Academy of Sciences.
P.~Z. is partially supported by the Innovation Program for Quantum Science and Technology project 2021ZD0301900.

\bibliography{refs}

\clearpage
\onecolumngrid
\renewcommand{\theequation}{S\arabic{equation}}
\setcounter{equation}{0} 

\renewcommand{\thefigure}{S\arabic{figure}}
\setcounter{figure}{0} 
\renewcommand{\thetable}{S\arabic{table}}
\setcounter{table}{0}

\section{Supplementary Material: Free-Energy Machine for Combinatorial Optimization}

\subsection{Derivation of variational mean-field free energy formulations for the benchmarking problems}

According to the definition of variational mean-field free energy, namely,
\begin{align}
F_{\rm MF}(\{P_i(\sigma_i)\},\beta) &= U_{\rm MF} - \frac{1}{\beta}S_{\rm MF} \;,
\end{align}
since the entropy term is irrelevant to $E(\bm{\sigma})$, we have the same entropy form for all benchmarking problems, as
\begin{gather}
S_{\rm MF} = -\sum_i\sum_{\sigma_i}P_i(\sigma_i)\ln P_i(\sigma_i) \; ,
\end{gather}
and it suffices to derive the formulae for different mean-field internal energies $U_{\rm MF}$.

For the maximum cut (MaxCut) problem, the energy function can be defined as the negation of the cut value,
\begin{align}
E(\bm{\sigma}) = -\sum_{(i,j)\,\in\,\mathcal{E}}W_{ij}[1-\delta(\sigma_i,\sigma_j)] \;,
\end{align}
where $\mathcal{E}$ is the edge set of the graph, $W_{ij}$ is the weight of edge $(i,j)$, and $\delta(\sigma_i,\sigma_j)$ stands for the delta function, which takes value 1 if $\sigma_i=\sigma_j$ and 0 otherwise.
Then the corresponding mean-field internal energy reads
\begin{align}
U^{\rm MaxCut}_{\rm MF}
&=\sum_{\bm{\sigma}}\prod_iP_i(\sigma_i)\left\{-\sum_{(i,j)\,\in\,\mathcal{E}}W_{ij}[1-\delta(\sigma_i,\sigma_j)]\right\}\\
\label{seq:free_energy_maxcut_3}
&=-\sum_{(i,j)\,\in\,\mathcal{E}}\sum_{\sigma_i,\sigma_j}W_{ij}P_i(\sigma_i)P_j(\sigma_j)[1-\delta(\sigma_i,\sigma_j)] \\
\label{seq:free_energy_maxcut_4}
&=-\sum_{(i,j)\,\in\,\mathcal {E}}\sum_{\sigma_i}W_{ij}P_i(\sigma_i)[1-P_j(\sigma_i)] \; ,
\end{align}
where identity $\sum_{\sigma_j}P_j(\sigma_j)\delta(\sigma_i,\sigma_j)=P_j(\sigma_i)$ is used in Eq.~(\ref{seq:free_energy_maxcut_3})-Eq.~(\ref{seq:free_energy_maxcut_4}). Thus, we have
\begin{align}
    F^{\rm MaxCut}_{\rm MF}(\{P_i(\sigma_i)\},\beta) &= -\sum_{(i,j)\,\in\,\mathcal {E}}\sum_{\sigma_i}W_{ij}P_i(\sigma_i)[1-P_j(\sigma_i)] \nonumber \\
    &+\frac{1}{\beta}\sum_i\sum_{\sigma_i}P_i(\sigma_i)\ln P_i(\sigma_i).
\end{align}

The energy function designed for the bMinCut problem given in the main text is 
\begin{align}
E(\bm{\sigma}) = \sum_{(i,j)\,\in\,\mathcal{E}}W_{ij}[1-\delta(\sigma_i,\sigma_j)] + \lambda\sum_i\sum_{j\neq i} \delta(\sigma_i, \sigma_j) \; , 
\end{align}
and the corresponding mean field internal energy reads
\begin{align}
U^{\rm bMinCut}_{\rm MF} &=\sum_{\bm{\sigma}}\prod_iP_i(\sigma_i)\left\{\sum_{(i,j)\,\in\,\mathcal{E}}W_{ij}[1-\delta(\sigma_i,\sigma_j)] + \lambda\sum_i\sum_{j\neq i} \delta(\sigma_i, \sigma_j)\right\} \\
&=\sum_{(i,j)\,\in\,\mathcal{E}}\sum_{\sigma_i,\sigma_j}W_{ij}P_i(\sigma_i)P_j(\sigma_j)[1-\delta(\sigma_i,\sigma_j)]+\lambda\sum_{i,j\neq i}\sum_{\sigma_i,\sigma_j}P_i(\sigma_i)P_j(\sigma_j)\delta(\sigma_i,\sigma_j)\\ 
&=\sum_{(i,j)\,\in\,\mathcal{E}}\sum_{\sigma_i}W_{ij}P_i(\sigma_i)[1-P_j(\sigma_i)] +  \lambda\sum_{i,j}\sum_{\sigma_i}[P_i(\sigma_i)P_j(\sigma_i)-P_i^2(\sigma_i)] \; .
\end{align}
Thus, we have
\begin{align}
    F^{\rm bMinCut}_{\rm MF}(\{P_i(\sigma_i)\},\beta) &= \sum_{(i,j)\, \in\,\mathcal E}\sum_{\sigma_i}W_{ij}P_i(\sigma_i)[1-P_j(\sigma_i)]\nonumber\\ 
    &+ \lambda\sum_{i,j}\sum_{\sigma_i}[P_i(\sigma_i)P_j(\sigma_i)-P^2_i(\sigma_i)]\nonumber\\
    &+\frac{1}{\beta}\sum_i\sum_{\sigma_i}P_i(\sigma_i)\ln P_i(\sigma_i).
\end{align}
The cost function designed for the Max-SAT problem given in the main text reads 
\begin{align}
E(\bm{\sigma}) = \sum_{m=1}^M\prod_{i\in\partial m}[1-\delta (W_{mi},\sigma_i)] \; .
\end{align}
The corresponding mean field internal energy is
\begin{align}
U^{\rm MaxSAT}_{\rm MF} &=\sum_{\bm{\sigma}}\prod_i P_i(\sigma_i)\left\{\sum_{m=1}^M\prod_{i\in\partial m}[1-\delta (W_{mi},\sigma_i)]\right\}\\
&=\sum_{m=1}^M\prod_{i\in\partial m}\sum_{\sigma_i}P_i(\sigma_i) [1-\delta (W_{mi},\sigma_i)]\\
&=\sum_{m=1}^M\prod_{i\in\partial m}[1-P_i(W_{mi})] \; .
\end{align}
Thus, we have
\begin{align}
\label{eq:maxksat_F}
    F^{{\rm MaxSAT}}_{\rm MF}(\{P_i(\sigma_i)\},\beta) &= \sum_{m=1}^M\prod_{i\in\partial m}\left[1-P_i(W_{mi})\right]\nonumber\\ 
   &+\frac{1}{\beta}\sum_i\sum_{\sigma_i}P_i(\sigma_i)\ln P_i(\sigma_i). 
\end{align}

\subsection{The Pytorch implementation of FEM using the automatic differentiation}
The following shows the Pytorch code with detailed annotations of implementing of FEM, on the MaxCut and bMinCut problems,
using the automatic differentiation.
Remarkably, even with the substantial disparities in the number of variable states, the existence of constraints, and the characteristics of the objective function between these two problems, the coding implementations for each only vary by a mere line.
\par

\begin{lstlisting}[language=Python]
import torch

def read_graph(file, index_start=0):
    """function for reading graph files
    the specific format should be n m in the first line, and m following lines
    represent source end weight
    Parameters:
        file: string, the filename of the graph to be read
        index_start: int, specify which is the start index of the graph"""
    with open(file,"r") as f:
        l = f.readline()
        n, m = [int(x) for x in l.split(" ") if x!="\n"]
        W = torch.zeros([n, n])
        for k in range(m):
            l = f.readline()
            l_split = l.split()
            i, j = [int(x) for x in l_split[:2]]
            if len(l_split) == 2:
                w = 1.0 # if no weight, 1.0 by default
            elif len(l_split) == 3:
                w = float(l_split[2])
            else:
                raise ValueError("Unkown graph file format")
            W[i-index_start, j-index_start], W[j-index_start, i-index_start] = w, w
    return n, m, W
    
def cut(W, p):
    """Calculating the cut size given the weight matrix and the marginal matrix"""
    p = p.to(W.dtype)
    return ((W @ p) * (1-p)).sum((1, 2))
    
def S(p):
    """Calculating the entropy of the marginal matrix p"""
    return -(p*p.log()).sum(2).sum(1)
    
def infer_cut(W, p):
    """Infer the cut size given the weight matrix and the marginal matrix"""
    s = torch.nn.functional.one_hot(
    p.argmax(dim=2), num_classes=p.shape[2])
    return s, cut(W, s) / 2

def balance(p):
    """Calculate the imbalance penalty of the marginal matrix p to restrict the group partition to be balanced"""
    return (p.sum(1)**2).sum(1)-(p**2).sum(2).sum(1)
    
def solve(problem, W, batch, q, penalty, beta_range, seed=0):
    """Function for solving the combinatorial optimization problem using FEM"""
    torch.manual_seed(seed)
    n = W.shape[0]
    h = torch.rand(batch, n, q) # field matrix with shape [batch, num_vertices, num_classes]
    h.requires_grad = True
    optimizer = torch.optim.Adam([h], lr=0.01)
    for beta in beta_range:
        p = torch.softmax(h, dim=2) # normalize the field matrix to get the marginal matrix
        if problem == 'maxcut':
            F = -cut(W,p) - S(p)/beta
        if problem == 'bmincut':
            F = cut(W,p)+penalty*balance(p)-S(p)/beta
        optimizer.zero_grad()
        F.backward(gradient=torch.ones_like(F))
        optimizer.step()
    return infer_cut(W, p)

n, m, W = read_graph('G1', index_start=1) # the data file can be retrieved from https://web.stanford.edu/~yyye/yyye/Gset/
beta_range = 1 / torch.linspace(10.0, 0.01, 1100)
configs, cuts = solve('maxcut', W, 1000, 2, 5.0, beta_range)
ind = torch.argmax(cuts)
print(configs[ind][:, 0], cuts[ind])

configs, cuts = solve('bmincut', W, 1000, 4, 5.0, beta_range)
ind = torch.argmax(cuts)
print(configs[ind], cuts[ind])
\end{lstlisting}

\subsection{The explicit gradient formulations}
The key step in FEM involves computing the gradients of $F_{\rm MF}$ with respect to the local fields $\{h_i(\sigma_i)\}$, denoted as $\{g^h_i(\sigma_i)\}$. This task can be accomplished by leveraging the automatic differentiation techniques. Additionally, we have the option to write down the explicit gradient formula for each problem at hand.   Once the specific form of $E(\bm{\sigma})$ is known, the form of $F_{\rm MF}$ is also determined. Thanks to the mean-field ansatz, the explicit formula for the gradients of $F_{\rm MF}$ with respect to $\{P_i(\sigma_i)\}$ can be obtained, denoted as $\{g^p_i(\sigma_i)\}$. The benefits of obtaining $\{g^p_i(\sigma_i)\}$ are twofold. 
Firstly, explicit gradient computation can lead to substantial time savings by eliminating the need for forward propagation calculations. Our numerical experiments have revealed that the application of explicit gradient formulations can reduce computational time by half.  Secondly, it enables problem-dependent gradient manipulations based on $\{g^p_i(\sigma_i)\}$, denoted as $\{\hat{g}^p_i(\sigma_i)\}$, enhancing numerical stability and facilitating smoother optimization within the gradient descent framework, extending beyond the conventional use of adaptive learning rates and momentum techniques commonly found in gradient-based optimization methods within the realm of machine learning.

Hence, in this work, we mainly adopt the explicit gradient approach for benchmarking FEM, and we use the manipulated gradients $\{\hat{g}^p_i(\sigma_i)\}$ to compute $\{g^h_i(\sigma_i)\}$. According to the chain rule for partial derivative calculation, $\{g^h_i(\sigma_i)\}$ can be computed from 
\begin{gather}
\label{eq:chain_rule}
    g^h_i(\sigma_i) = \sum_{\sigma_i^\prime}\frac{\partial F_{\rm MF}}{\partial P_i(\sigma_i^\prime)}\frac{\partial P_i(\sigma_i^\prime)}{\partial h_i(\sigma_i)} \; ,
\end{gather}
and since $P_i(\sigma_i) = e^{h_i(\sigma_i)}/Z$, where $Z=\sum_{\sigma_i^\prime}e^{h_i(\sigma_i^\prime)}$ is the normalization factor, we also have
\begin{align}
\label{eq:paticial_deri_p_h}
\begin{split}
\frac{\partial P_i(\sigma_i^\prime)}{\partial h_i(\sigma_i)}= \left \{
\begin{array}{ll}
-\frac{-e^{h_i(\sigma_i^\prime)}}{Z}\frac{e^{h_i(\sigma_i)}}{Z}=-P_i(\sigma_i^\prime)P_i(\sigma_i),   & \sigma_i^\prime \neq \sigma_i\\[6pt]
\frac{e^{h_i(\sigma_i)}}{Z}(1-\frac{e^{h_i(\sigma_i)}}{Z}) = P_i(\sigma_i)[1-P_i(\sigma_i)],     & \sigma_i^\prime = \sigma_i \; .
\end{array}
\right.
\end{split}
\end{align}
Therefore, by substituting Eq.~(\ref{eq:paticial_deri_p_h}) into Eq.~(\ref{eq:chain_rule}) and employing the modified gradient variable $\hat{g}^p_i(\sigma_i^\prime)$ in place of $\frac{\partial F_{\rm MF}}{\partial P_i(\sigma_i^\prime)}$ (instead of using $g^p_i(\sigma_i)$ directly), we have the following unified form for $\{g^h_i(\sigma_i)\}$, 
\begin{gather}
    \label{eq:gradients_wrt_h}
    g^h_i(\sigma_i) = \gamma_{\rm grad} \left[\hat{g}^p_i(\sigma_i) -\sum_{\sigma_i^\prime=1}^qP_i(\sigma_i^\prime)\hat{g}^p_i(\sigma_i^\prime)\right]P_i(\sigma_i) \;,
\end{gather}
where $\gamma_{\rm grad}$  is a hyperparameter that controls the magnitudes of $\{g^h_i(\sigma_i)\}$ to accommodate different optimizers. Once we have the values of $\{P_i(\sigma_i)\}$ and $\{\hat{g}^p_i(\sigma_i)\}$ (also computed using $\{P_i(\sigma_i)\}$), we can obtain $\{g^h_i(\sigma_i)\}$  immediately according to Eq.~(\ref{eq:gradients_wrt_h}).

\subsection{The explicit gradients and manipulated gradients for the benchmarking problems}
\subsubsection{The MaxCut problem} 
The explicit gradients of $\{g^p_i(\sigma_i)\}$ for the MaxCut problem can be derived analytically from $F_{\rm MF}^{\rm MaxCut}$, as
\begin{align}
\label{seq:free_energy_maxcut_gradient}
g_i^p(\sigma_i)=\nabla_{P_i(\sigma_i)}F_{\rm MF}^{\rm MaxCut}=\sum_jW_{ij}[2P_j(\sigma_i)-1] + \frac{1}{\beta}[\ln P_i(\sigma_i) + 1] \; .
\end{align}

The other derivation method may be based on by rewritting Eq.~(\ref{seq:free_energy_maxcut_4}) as 
\begin{align}
U^{\rm MaxCut}_{\rm MF}
&=-\sum_{(i,j)\,\in\,\mathcal {E}}\sum_{\sigma_i}W_{ij}P_i(\sigma_i)[1-P_j(\sigma_i)] \\
&=-\sum_{(i,j)\,\in\,\mathcal {E}}W_{ij}+\sum_{(i,j)\,\in\,\mathcal {E}}W_{ij}P_i(\sigma_i)P_j(\sigma_i) \;,   
\end{align}
where the first term $-\sum_{(i,j)\,\in\,\mathcal {E}}W_{ij}$ is a constant. Hence, we also have the following different formula
\begin{align}
\label{seq:free_energy_maxcut_gradient_2}
g_i^p(\sigma_i)=2\sum_jW_{ij}P_j(\sigma_i) + \frac{1}{\beta}[\ln P_i(\sigma_i) + 1] \; .
\end{align}
However, using Eq.~(\ref{seq:free_energy_maxcut_gradient}) or Eq.~(\ref{seq:free_energy_maxcut_gradient_2}) will result in the same result for computing $g_i^h(\sigma_i)$, for the reason that the constant $-\sum_{j}W_{ij}$ in Eq.~(\ref{seq:free_energy_maxcut_gradient}) for each index $i$ will be canceled when computing the gradients for the local fields using Eq.~(\ref{eq:gradients_wrt_h}).
Then the manipulated gradients can be designed as follows
\begin{gather}
\label{eq:maxcut_modified_gradient}
    \hat{g}^p_i(\sigma_i) = c_i\sum_jW_{ij}e_{j}(\sigma_i) + \frac{1}{\beta}[\ln P_i(\sigma_i) + 1] \; ,
\end{gather}
where two modifications on gradient have been made to improve the optimization performance. Firstly, $[e_{j}(1),e_{j}(2), \ldots, e_{j}(q)]$ is the one-hot vector (length of 2 in MaxCut problem) corresponding to $[P_{j}(1),P_{j}(2), \ldots, P_{j}(q)]$. The introduction of $\{e_{j}(\sigma_j)\}$ to replace $\{P_{j}(\sigma_j)\}$ is called the discretization that enables reducing the analog errors introduced by $\sum_jW_{ij}P_{j}(\sigma_j)$ in the explicit gradients. Note that the similar numerical tricks have been also employed in the previous work~\cite{goto2021high}. Secondly, when the graph has inhomogeneity in the node degrees or the edge weights, the magnitude of $\sum_jW_{ij}e_{j}(\sigma_j)$ for each spin variable can exhibit significant differences. Hence, the gradient normalization factor $c_i$ enables robust optimization and better numerical performance. 

For the MaxCut problem, the gradient normalization factor $c_i$ is set to $c_i=\frac{1}{\sum_j|W_{ij}|}$ in this work. In this context, the $L_1$ norm, represented by $\sum_j|W_{ij}|$, normalizes the gradient magnitude for each spin, ensuring that the values of $\sum_jW_{ij}e_j(\sigma_j)$ across all spins remain below one. This normalization is critical. Since the range of the gradients of the entropy term, given by $\ln P_i(\sigma_i) + 1$, is consistent across all spins. In our experiments, we found the constrains made on the gradient magnitude of the internal energy can prevent the system from becoming ensnared in local minima. Although other tricks for the normalization factors can be employed, we have observed that our current settings yield satisfactory performance in our numerical experiments.
\vspace{1.1em}

\subsubsection{The bMinCut problem}
In the bMinCut problem, the gradients with respect to $P_i(\sigma_i)$ are
\begin{align}
g^p_i(\sigma_i)=\nabla_{P_i(\sigma_i)}F_{\rm MF}^{\rm bMinCut} = -2\sum_jW_{ij}P_j(\sigma_i)+2\lambda[\sum_jP_j(\sigma_i)-P_i(\sigma_i)] + \frac{1}{\beta}[\ln P_i(\sigma_i) + 1]\; .
\end{align}
The modifications can be made for better optimization on graphs with different topologies, as done in the MaxCut problem. We have the following manipulated gradients
\begin{gather}
\hat{g}_i^p(\sigma_i)= -c_i^f\sum_jW_{ij}e_j(\sigma_i)+\lambda c_i^a[\sum_je_j(\sigma_i)-e_i(\sigma_i)] + \frac{1}{\beta}[\ln P_i(\sigma_i) + 1] \; ,
\end{gather}
where $c_i^f$ and $c_i^a$ are the gradient normalization factors for the ferromagnetic and antiferromagnetic terms, respectively. The $\{e_i(\sigma_i)\}$ are again the one-hot vectors for $\{P_i(\sigma_i)\}$, which serves to mitigate the analog error introduced in the explicit gradients, as we done in the MaxCut problem. 

For the bMinCut problem, analogous to the approach taken with the MaxCut problem, the ferromagnetic normalization factor $c_i^f$ is defined as $c_i^f=\frac{q}{\sum_j W_{ij}}$ (in the bMinCut problem, $W_{ij}>0$), while the antiferromagnetic normalization factor $c_i^a$ is determined by $c_i^a=\frac{q}{\sqrt{\sum_jW_{ij}}}$. The rationale behind the setting for $c_i^a$ stems from the intuition that spins with larger values of $\sum_j W_{ij}$ should remain in their current states, implying that they should not be significantly influenced by the antiferromagnetic force to transition into other states. Although theses settings for the normalization factors may not be optimal, and alternative schemes could be implemented, we have observed that these particular settings yield satisfactory performance in our numerical experiments.
\vspace{1.1em}

\subsubsection{The MaxSAT problem}
The explicit gradient can be readily computed as follows
\begin{align}
g_i^p(\sigma_i)=-\sum_{m=1}^M\delta(\sigma_i,W_{mi})\prod_{k\in\partial m,\ k\neq i}[1-P_k(W_{kj})] + \frac{1}{\beta}[\ln P_i(\sigma_i) + 1] \; .
\end{align}
Note that, for each $m$ in the summation, the spin variable $\sigma_i$ must appear as a literal in the clasue $C_m$, otherwise the gradients of the internal energy with respect to $\sigma_i$ are zeros in this clause. For the random Max $k$-SAT problem benchmarked in this work, we make no modifications. Therefore, we set $\hat{g}_i^p(\sigma_i)$ equal to $g_i^p(\sigma_i)$.

\subsection{Numerical experiments on the MaxCut problem}

\subsubsection{Simplification of FEM for solving the MaxCut problem} 
To facilitate an efficient implementation, we can simplify FEM's approach for solving the Ising problem since the gradients for $\{h_i(+1)\}$ and $\{h_i(-1)\}$ are dependent. 
In the MaxCut problem with $q=2$ ($\sigma_i=+1$ or $\sigma_i=-1$), it is straightforward to prove $g^h_i(+1)=-g^h_i(-1)=\gamma_{\rm grad}[\hat{g}^p_i(+1)-\hat{g}^p_i(-1)]P_i(+1)P_i(-1)$ from Eq.~(\ref{eq:gradients_wrt_h}). 
Given that $P_i(+1)=1-P_i(-1)=e^{h_i(+1)}/(e^{h_i(+1)}+e^{h_i(-1)})={\rm sigmoid} (h_i(+1)-h_i(-1))$, we actually only need to update $\{P_i(+1)\}$ (or simply $\{P_i\}$) to save considerable computational resources. Thus, we introduce the new local field variables $\{h_i\}$ to replace $\{h_i(+1)-h_i(-1)\}$, such that $P_i={\rm sigmoid}(h_i)$. 

According to Eq.~(\ref{eq:maxcut_modified_gradient}), $\hat{g}^p_i(+1)=c_i\sum_jW_{ij}e_j(+1)+\frac{1}{\beta}[\ln P_i(+1) + 1]$ and $\hat{g}^p_i(-1)=c_i\sum_jW_{ij}e_j(-1)+\frac{1}{\beta}[\ln P_i(-1) + 1]$. Based on $g^h_i(+1)=\gamma_{\rm grad}[\hat{g}^p_i(+1)-\hat{g}^p_i(-1)]P_i(+1)P_i(-1)$, the gradients regarding to local fields $\{h_i\}$ can be written as 
\begin{gather}
    g^h_i = \gamma_{\rm grad}[\hat{g}^p_i(+1)-\hat{g}^p_i(-1)]P_i(+1)P_i(-1) \nonumber\\
    \label{eq:grad_single_h2}
    =\gamma_{\rm grad}\left[c_i\sum_jW_{ij}(2e_j(+1)-1)+\frac{1}{\beta}\ln \frac{P_i}{1-P_i}\right]P_i(1-P_i) \; .
\end{gather}
We can further simplify Eq.~(\ref{eq:grad_single_h2}) by introducing the magnetization $m_i=2P_i-1=\tanh(h_i/2)$, such that 
\begin{gather}
\label{eq:grad_m_h}
    g^h_i = \frac{\gamma_{\rm grad}}{4}[c_i\sum_jW_{ij}{\rm sgn}(m_j)+\frac{1}{\beta}h_i](1-m_i^2) \; ,
\end{gather}
where ${\rm sgn}(\cdot)$ is the sign function, and the identity ${\rm arctanh}(x)=\frac{1}{2}\ln \frac{1+x}{1-x}$ has been used for the simplifications. Thus,  we have used $\{m_i\}$ and $\{h_i\}$ to simplify the original gradients, requiring optimizations for only half of the variational variables as compared to the non-simplified case.

\subsubsection{The numerical experiment of the MaxCut problem on the complete graph $K_{2000}$}
For the numerical experiment of the MaxCut problem on the complete graph $K_{2000}$, as shown in Fig. 3(a) in the main text (where we evaluate the performance by only varying the total annealing steps $N_{\rm step}$ while keeping other hyperparameters unchanged), we implemented the dSBM algorithm, which is about to simulate the following Hamiltonian equations of motion~\cite{goto2021high}
\begin{gather}
    y_i(t_{k+1}) =  y_i(t_k) + \left\{-[a_0-a(t_k)]x_i(t_k) + c_0\sum_{j=1}^NW_{ij}{\rm sgn} [x_j(t_k)] \right \} \Delta_t \; ,\\
    x_i(t_{k+1}) = x_i(t_k) + a_0y_i(t_{k+1})\Delta_t \; ,
\end{gather}
where $x_i$ and $y_i$ represent the position and momentum of a particle corresponding to the $i$-th spin in a $N$-particle dynamical system, respectively, $\Delta_t$ is the time step, $t_k$ is discrete time with $t_{k+1} = t_k+\Delta_t$, $J$ is the edge weight matrix, $a(t_k)$ is the bifurcation parameter linearly increased from 0 to $a_0=1$, and $c_0=0.5\sqrt{\frac{N-1}{\sum_{i,j}W_{ij}^2}}$ is according to the settings in Ref.~\cite{goto2021high}. In addition, at every $t_k$, if $|x_i|>1$, we set $x_i={\rm sgn}(x_i)$ and $y_i=0$. For the dSBM benchmarks on $K_{2000}$, we set $\Delta_t=1.25$ following the recommended settings in Ref.~\cite{goto2021high}, and the initial values of $x_i$ and $y_i$ are randomly initialized from the range $[-0.1, 0.1]$. Regarding FEM, we employ the explicit gradient formulations, setting the values of $\gamma_{\rm grad}=1$, $T_{\rm max}$ = 1.16, $T_{\rm min}$ = 6e-5, and utilize the inverse-proportional scheduling for annealing. We employ RMSprop as the optimizer, with the optimizer hyperparameters alpha, momentum, weight decay, and learning rate set to 0.56, 0.63, 0.013, and 0.03, respectively. Both dSBM and FEM were executed on a GPU.

\subsubsection{The numerical experiment of the MaxCut problem on the G-set}
For the benchmarks on the G-set problems, we have presented the detailed TTS results obtained by FEM in Tab.~\ref{tab:gset_TTS}, along with a comparison of the reported data provided by dSBM. Given the capability of FEM for optimizing many replicas parallelly, here we assess the TTS using the batch processing method introduced in Ref.~\cite{goto2021high}. All the parameter settings for FEM are listed in Tab.~\ref{tab:gset_param}. We also utilized the same GPU used in Ref.~\cite{goto2021high} for implementing FEM in this benchmark. In this study, we adopt the batch processing method as introduced in Ref.~\cite{goto2021high} for calculating TTS. Therefore, for an accurate comparison, the values of $N_{rep}$ for each instance shown in Tab.~\ref{tab:gset_param} are consistent with those used for dSBM in Ref.~\cite{goto2021high}. Throughout the benchmarking, we initialize the local fields with random values according to $h_i(\sigma_i)^{ini} = 0.001 * {\rm randn}$, where ${\rm randn}$ represents a random number sampled from the standard Gaussian distribution. All variables are represented using 32-bit single-precision floating-point numbers.

\setlength{\tabcolsep}{1mm}
\setlength{\LTcapwidth}{0.95\textwidth}
{\footnotesize
\begin{longtable}[c]{c|ccc|ccc|cccccc}
\caption{\textbf{Benchmarking results on the G-Set instances}. The TTS is defined as the time taken to achieve the best cut found by the solver, and the parentheses for the TTS results indicate that the best cut for computing TTS is not the best known cut. Shorter TTS results are highlighted in bold.\label{tab:gset_TTS}}\\
\Xcline{1-10}{0.8pt}
 \multirow{2}*{\makecell[c]{\textbf{Graph}\\\textbf{type}}}&\multirow{2}*{\textbf{Instance}} & \multirow{2}*{\textit{\textbf{N}}} & \multirow{2}*{\makecell[c]{\textbf{Best}\\\textbf{known}}} & \multicolumn{3}{c|}{\textbf{FEM}} &\multicolumn{3}{c}{\textbf{dSBM}}\\
 ~ & ~&~ & ~ & \textbf{Best cut} & \textbf{TTS(ms)} & \boldmath{$P_s$} & \textbf{Best cut} & \textbf{TTS(ms)} & \boldmath{$P_s$}\\
\Xcline{1-10}{0.8pt}
\endfirsthead
\Xcline{1-10}{0.8pt}
 \multirow{2}*{\makecell[c]{\textbf{Graph}\\\textbf{type}}}&\multirow{2}*{\textbf{Instance}} & \multirow{2}*{\textit{\textbf{N}}} & \multirow{2}*{\makecell[c]{\textbf{Best}\\\textbf{known}}} & \multicolumn{3}{c|}{\textbf{FEM}} &\multicolumn{3}{c}{\textbf{dSBM}}\\
 ~ & ~&~ & ~ & \textbf{Best cut} & \textbf{TTS(ms)} & \boldmath{$P_s$} & \textbf{Best cut} & \textbf{TTS(ms)} & \boldmath{$P_s$}\\
\Xcline{1-10}{0.8pt}
\endhead
\Xcline{1-10}{0.8pt}
\multicolumn{12}{c}{\textbf{\normalsize to be continued...}}
\endfoot
\Xcline{1-10}{0.8pt}
\endlastfoot

~& G1 & 800 & 11624 & 11624 & \textbf{24.9} & 100.0\% &11624& 33.3 & 98.7\% \\
~& G2 & 800 & 11620 & 11620 & \textbf{96.5} & 99.6\% &11620& 239 & 82\% \\
~& G3 & 800 & 11622 & 11622 &\textbf{23.3} & 100.0\% &11622& 46.2 & 99.6\% \\
\multirow{5}*{Random}& G4 & 800 & 11646 & 11646 & \textbf{19.8} & 99.5\% &11646& 34.4 & 98.3\%\\
~& G5 & 800 & 11631 & 11631 & \textbf{21.6} & 98.9\% &11631& 58.6 & 97.2\%\\
~&G6 & 800 & 2178 & 2178 & \textbf{5.6} & 95.5\% & 2178 & 6.3 & 97.9\%\\
~&G7 & 800 & 2006 & 2006 &  11.5 & 98.6\% & 2006 & \textbf{6.85} & 97.4\%\\
~&G8 & 800 & 2005 & 2005 & 21.3 & 98.5\% & 2005 &\textbf{11.9} & 95.4\%\\
~&G9 & 800 & 2054 & 2054 & \textbf{35.6} & 98.8\% & 2054 & 36 & 86.7\%\\
~&G10 & 800 & 2000 & 2000 & 193 & 53.2\% & 2000 & \textbf{47.7} & 40.7\% \\
\Xcline{1-10}{0.5pt}
\multirow{3}*{Toroidal}& G11 & 800 & 564 & 564 & 24.2 & 99.0\% & 564 &\textbf{3.49} & 98\% \\
~&G12 & 800 & 556 & 556 & 31.2 & 97.8\% & 556 &\textbf{5.16} & 97.3\% \\
~&G13 & 800 & 582 & 582 & 203 & 63.9\% & 582 &\textbf{11.9}& 99.6\% \\
\Xcline{1-10}{0.5pt}
\multirow{8}*{Planar}&G14 & 800 & 3064 & 3064 & \textbf{2689}& 36.5\% & 3064 & 71633 & 0.5\%\\
~&G15 & 800 & 3050 & 3050 &\textbf{164}& 96.5\% & 3050 &340 & 80.4\% \\
~&G16 & 800 & 3052 & 3052 & \textbf{165}& 99.8\% & 3052 &347 & 99.2\%\\
~&G17 & 800 & 3047 & 3047 & \textbf{800}& 70.2\%& 3047 &1631& 28.3\% \\
~&G18 & 800 & 992 & 992 & \textbf{264}& 37.9\% & 992 & 375 & 7.4\% \\
~&G19 & 800 & 906 & 906 & \textbf{17.5}& 98.8\% & 906 &17.8& 99.5\%\\
~&G20 & 800 & 941 & 941 & \textbf{8.5}& 99.2\%  & 941 &9.02& 98\%\\
~&G21 & 800 & 931 & 931 & \textbf{67.3}& 34\%& 931 &260 &13.6\% \\
\Xcline{1-10}{0.5pt}
\multirow{10}*{Random}& G22 & 2000 & 13359 & 13359 & 917 & 56.3\%  & 13359 &\textbf{429} &92.8\% \\
~&G23 & 2000 & 13344 & 13342 & (98) & 36.9\%  & 13342 &\textbf{(89)} & - \\
~&G24 & 2000 & 13337 & 13337 & 1262 & 92.4\% & 13337 &\textbf{459} & 64.8\% \\
~& G25 & 2000 & 13340 & 13340 & 5123 &31.9\% & 13340 &\textbf{2279} & 39.9\% \\
~&G26 & 2000 & 13328 & 13328 & 991 & 83.2\%  & 13328& \textbf{476}& 64.3\% \\
~&G27 & 2000 & 3341 & 3341 & 127 & 90.1\% & 3341& \textbf{49.9}& 97.1\% \\
~&G28 & 2000 & 3298 & 3298 & 306 & 82.7\%  & 3298& \textbf{87.2} & 95.2\% \\
~&G29 & 2000 & 3405 & 3405 & \textbf{200} & 98.6\% & 3405& 221& 73.7\% \\
~&G30 & 2000 & 3413 & 3413& 948 & 64.7\% & 3413 &\textbf{439}& 73.8\% \\
~&G31 & 2000 & 3310 & 3310 & 4523 & 19.6\% & 3310 &\textbf{1201} &19.9\%\\
\Xcline{1-10}{0.5pt}
\multirow{3}*{Toroidal}&G32 & 2000 & 1410 & 1410 & 23749 & 1.3\%& 1410 &\textbf{3622}& 9.3\%\\
~&G33 & 2000 & 1382 & 1382 & 659607 & 0.6\% & 1382 &\textbf{57766}& 0.5\%\\
~&G34 & 2000 & 1384 & 1384 & 12643 & 28.1\% & 1384 &\textbf{2057} & 23.1\%\\
\Xcline{1-10}{0.5pt}
\multirow{8}*{Planar} & G35 & 2000 & 7687 & 7686 &\textbf{ (5139390)} & 0.01\% & 7686 & (8319000)& - \\
~&G36 & 2000 & 7680 & 7680 & \textbf{5157009} & 0.01\%  & 7680 &62646570&  0.01\%\\
~&G37 & 2000 & 7691 & 7690 &\textbf{(3509541)} & 0.01\%  & 7691 &27343457&0.02\%\\
~&G38 & 2000 & 7688 & 7688 & \textbf{41116}& 7.3\%  & 7688& 98519&6.8\%\\
~&G39 & 2000 & 2408 & 2408 & \textbf{12461} & 17.5\%  & 2408 &56013 & 10.7\%\\
~&G40 & 2000 & 2400 & 2400 & \textbf{3313} & 54.1\%  & 2400 &24131 & 15.4\% \\
~&G41 & 2000 & 2405 & 2405 & \textbf{1921} & 80.9\%& 2405 &10585 &28.2\% \\
~&G42 & 2000 & 2481 & 2481 & \textbf{91405} & 0.23\% & 2480 &(550000)& - \\
\Xcline{1-10}{0.5pt}
\multirow{5}*{Random}&G43 & 1000 & 6660 & 6660 & 19.8 & 66.1\%  & 6660 &\textbf{5.86} &99.2\% \\
~&G44 & 1000 & 6650 & 6650 & 13.2 & 80.1\%  & 6650 &\textbf{6.5} &98.5\% \\
~&G45 & 1000 & 6654 & 6654 & \textbf{35} & 98.7\% & 6654 &43.4& 98.5\% \\
~&G46 & 1000 & 6649 & 6649 & 141 & 69.8\%  & 6649 &\textbf{16} &99.2\% \\
~&G47 & 1000 & 6657 & 6657 & \textbf{33.9}&98.7\% & 6657 &44.8 &98.2\% \\
\Xcline{1-10}{0.5pt}

~&G48 & 3000 & 6000 & 6000 & \textbf{0.35} & 95.3\% & 6000 &0.824 &100.0\% \\
\multirow{2}*{Toroidal}&G49 & 3000 & 6000 & 6000 & \textbf{0.66} & 97.8\% & 6000 &0.784 &99.5\% \\
~&G50 & 3000 & 5880 & 5880 & \textbf{2.01} & 85.1\%  & 5880 &2.63 &100.0\% \\
\Xcline{1-10}{0.5pt}
\multirow{4}*{Planar}&G51 & 1000 & 3848 & 3848 & \textbf{5268} & 16.1\% & 3848 &12209 &6.7\%\\
~&G52 & 1000 & 3851 & 3851 & \textbf{4580} & 30.4\% &3851 &6937 &21.3\% \\
~&G53 & 1000 & 3850 & 3850 & \textbf{6155} & 24.1\% &3850 &93899 &4.3\% \\
~&G54 & 1000 & 3852 & 3852 & \textbf{55055} & 0.24\%  & 3852 &2307235 &0.06\% \\
\Xcline{1-12}{0.8pt}

\end{longtable}
}

{\renewcommand{\arraystretch}{1.3}
\setlength{\tabcolsep}{1mm}
\setlength{\LTcapwidth}{1\textwidth}
{\footnotesize
\begin{longtable}[c]{ccccccccccccc}
\caption{\textbf{The hyperparameter settings for FEM in the TTS benchmarking on G-set instances.} Here, we utilize different optimizers, SGD and RMDprop, for different sets of instances in this benchmarking based on their performance. For the explanations of the hyperparameters of the different optimizers. The inverse-proportional scheduling is used for the annealing. The meanings of $N_{\rm step}$, $N_{\rm batch}$ and $N_{\rm rep}$ are the same with the TTS experiments documented in Ref.~\cite{goto2021high}. For $N_{\rm batch}$ in FEM, we refer to the number of replicas. \label{tab:gset_param}}\\
\Xcline{1-13}{0.9pt}
\multirow{2}*{\textbf{Ins.}} &\multirow{2}*{\boldmath{$T_{\rm max}$}}&\multirow{2}*{\boldmath{$T_{\rm min}$}}&\multirow{2}*{\boldmath{$\gamma_{\rm grad}$}}&\multirow{2}*{\textbf{Optimizer}}&\multirow{2}*{\textit{\textbf{lr}}}&\multirow{2}*{\textbf{alpha}}&\multirow{2}*{\makecell[c]{\textbf{dampe-}\\\textbf{ning}}}&\multirow{2}*{\makecell[c]{\textbf{weight}\\\textbf{decay}}}&\multirow{2}*{\makecell[c]{\textbf{momen-}\\\textbf{tum}}}&\multirow{2}*{\boldmath{$N_{\rm step}$}}&\multirow{2}*{\boldmath{$N_{\rm batch}$}}&\multirow{2}*{\boldmath{$N_{\rm rep}$}}\\
~&~&~&~&~&~&~&~&~&~&~&~&~\\
\Xcline{1-13}{0.9pt}
\endfirsthead

\Xcline{1-13}{0.9pt}
\multirow{2}*{\textbf{Ins.}} &\multirow{2}*{\boldmath{$T_{\rm max}$}}&\multirow{2}*{\boldmath{$T_{\rm min}$}}&\multirow{2}*{\boldmath{$\gamma_{\rm grad}$}}&\multirow{2}*{\textbf{Optimizer}}&\multirow{2}*{\textit{\textbf{lr}}}&\multirow{2}*{\textbf{alpha}}&\multirow{2}*{\makecell[c]{\textbf{dampe-}\\\textbf{ning}}}&\multirow{2}*{\makecell[c]{\textbf{weight}\\\textbf{decay}}}&\multirow{2}*{\makecell[c]{\textbf{dampe-}\\\textbf{ning}}}&\multirow{2}*{\boldmath{$N_{\rm step}$}}&\multirow{2}*{\boldmath{$N_{\rm batch}$}}&\multirow{2}*{\boldmath{$N_{\rm rep}$}}\\
~&~&~&~&~&~&~&~&~&~&~&~&~\\
\Xcline{1-13}{0.9pt}
\endhead

\multicolumn{13}{c}{\textbf{\normalsize to be continued...}}
\endfoot
\Xcline{1-13}{0.9pt}
\endlastfoot

G1 & 0.5 & 8e-5 & 1 & RMSprop & 0.2 & 0.623 &-&0.02&0.693&1000&130&1000\\
\Xcline{1-13}{0.5pt}
G2 & 0.2592 & 6.34e-4& 1 & RMSprop & 0.0717 & 0.5485 & - & 0.0264&0.9082& 5000 & 100 & 1000\\
\Xcline{1-13}{0.5pt}
G3 & 0.264 & 1.1e-3& 1 & RMSprop & 0.3174 & 0.7765 & - & 0.00672&0.7804& 1000 & 120 & 1000\\
\Xcline{1-13}{0.5pt}
G4 & 0.29 & 8.9e-4& 1 & RMSprop & 0.2691 & 0.4718 & - & 0.00616&0.7414& 800 & 130 & 1000\\
\Xcline{1-13}{0.5pt}
G5 & 0.2 & 9e-4& 1 & RMSprop & 0.24 & 0.9999 & - & 0.0056&0.8215& 1000 & 110 & 1000\\
\Xcline{1-13}{0.5pt}
G6 & 0.44 & 1.7e-3& 1 & RMSprop & 0.534 & 0.6045 & - & 0.00657&0.4733& 1000 & 20 & 1000\\
\Xcline{1-13}{0.5pt}
G7& 0.54 & 1.8e-3& 1 & RMSprop & 0.452 & 0.8966 & - & 0.0087&0.632& 700 & 80 & 1000\\
\Xcline{1-13}{0.5pt}
G8 & 0.19 & 7.92e-4& 1 & RMSprop & 0.296 & 0.9999 & - & 0.00731&0.737& 1000 & 100 & 1000\\
\Xcline{1-13}{0.5pt}
G9 & 0.208 & 9e-4& 1 & RMSprop & 0.305 & 0.9999 & - & 0.00205&0.718& 2500 & 70 & 1000\\
\Xcline{1-13}{0.5pt}
G10 & 1.28 & 5.21e-6& 0.75 & SGD & 1.2 & - & 0.082 & 0.03&0.88& 2000 & 100 & 1000\\
\Xcline{1-13}{0.5pt}
G11 & 1.28 & 4.96e-6& 0.98 & SGD & 1.2 & - & 0.13 & 0.061&0.88& 1800 & 120 & 1000\\
\Xcline{1-13}{0.5pt}
G12 & 1.28 & 7.8e-6& 0.65 & SGD & 1.98 & - & 0.13 & 0.06&0.88& 1600 & 140 & 1000\\
\Xcline{1-13}{0.5pt}
G13 & 1.28 & 3.12e-6& 1.7 & SGD & 3 & - & 0.082 & 0.033 &0.76& 3000 & 130 & 1000\\
\Xcline{1-13}{0.5pt}
G14 & 0.387 & 8.64e-4& 1 & RMSprop & 0.44 & 0.9999 & - & 0.0089 &0.793& 7000 & 250 & 1000\\
\Xcline{1-13}{0.5pt}
G15& 0.5 & 1e-3& 1 & RMSprop & 0.45 & 0.9999 & - & 0.0056 &0.7327& 4000 & 200 & 1000\\
\Xcline{1-13}{0.5pt}
G16 & 0.54 & 8.1e-4& 1 & RMSprop & 0.288 & 0.9999 & - & 0.00756 &0.7877& 7000 & 160 & 1000\\
\Xcline{1-13}{0.5pt}
G17  & 0.253& 1.06e-3& 1 & RMSprop & 0.631 & 0.9999 & - & 0.01341 &0.7642& 7000 & 200 & 1000\\
\Xcline{1-13}{0.5pt}
G18  & 0.4 & 1e-3& 1 & RMSprop & 0.345 & 0.99 & - & 0.01 &0.9& 1200 & 150 & 1000\\
\Xcline{1-13}{0.5pt}
G19  & 0.962 & 3.98e-6& 1.75 & SGD & 4.368 & - & 0.05175 & 0.01336 &0.729& 1700 & 85 & 1000\\
\Xcline{1-13}{0.5pt}
G20  & 0.37 & 9.4e-4& 1.55 & RMSprop & 1.38 & 0.9089 & - & 0.00445 & 0.8186 & 500 & 100 & 1000\\
\Xcline{1-13}{0.5pt}
G21  & 0.6 & 9.6e-4& 1 & RMSprop & 0.33 & 0.9999 & - & 0.0092 & 0.692 & 1000 & 40 & 1000\\
\Xcline{1-13}{0.5pt}
G22  & 0.352 & 2.4e-4& 1 & RMSprop & 0.481 & 0.9999 & - & 0.00382 & 0.7166 & 4700 & 90 & 1000\\
\Xcline{1-13}{0.5pt}
G23  & 0.406 & 1.15e-6& 2.72 & SGD & 8.042 & - & 0.1443 & 0.00184 & 0.714 & 3200 & 10 & 1000\\
\Xcline{1-13}{0.5pt}
G24  & 0.528 & 1.6e-4& 1 & RMSprop & 0.39 & 0.9999 & - & 0.00413 & 0.74 & 7000 & 250 & 1000\\
\Xcline{1-13}{0.5pt}
G25  & 0.4 & 4.83e-6& 5.33 & SGD & 3.66 & - & 0.0905 & 0.00987 & 0.672 & 7000 & 200 & 1000\\
\Xcline{1-13}{0.5pt}
G26  & 0.361 & 4.43e-6& 2.18 & SGD & 8.46 & - & 0.0612 & 0.0078 & 0.714 & 6000 & 200 & 1000\\
\Xcline{1-13}{0.5pt}
G27 & 0.28 & 5e-4& 1 & RMSprop & 0.7 & 0.9995 & - & 0.00575 & 0.78 & 2000 & 80 & 1000\\
\Xcline{1-13}{0.5pt}
G28  & 0.32 & 5e-4& 1 & RMSprop & 0.69 & 0.999 & - & 0.006 & 0.78 & 3000 & 100 & 1000\\
\Xcline{1-13}{0.5pt}
G29 & 0.38 & 2.7e-4& 1 & RMSprop & 0.44 & 0.9999 & - & 0.013 & 0.7 & 4000 & 120 & 1000\\
\Xcline{1-13}{0.5pt}
G30 & 0.96 & 4.92e-6& 1.9 & SGD & 2.59 & - & 0.05 & 0.053 & 0.715 & 7000 & 100 & 1000\\
\Xcline{1-13}{0.5pt}
G31  & 1.834 & 2.76e-6& 1.32 & SGD & 1.38 & - & 0.0104 & 0.083 & 0.7566 & 7000 & 100 & 1000\\
\Xcline{1-13}{0.5pt}
G32  & 0.89 & 1.42e-5& 3.17 & SGD & 1.67 & - & 0.1285 & 0.018 & 0.9 & 12000 & 20 & 1000\\
\Xcline{1-13}{0.5pt}
G33 & 0.605 & 7.8e-6& 2 & SGD & 4.05 & - & 0.098 & 0.0366 & 0.91 & 12000 & 260 & 1000\\
\Xcline{1-13}{0.5pt}
G34 & 0.605 & 6.24e-6& 2.33 & SGD & 2.638 & - & 0.1182 & 0.0384 & 0.8967 & 12000 & 260 & 1000\\
\Xcline{1-13}{0.5pt}
G35 & 0.9 & 1e-4& 1 & RMSprop & 0.023 & 0.9999 & - & 0.016 & 0.92 & 15000 & 20 & 10000\\
\Xcline{1-13}{0.5pt}
G36 & 1 & 1e-3& 1 & RMSprop & 0.1 & 0.999 & - & 0.025 & 0.89 & 12000 & 25 & 10000\\
\Xcline{1-13}{0.5pt}
G37 & 0.9 & 1e-4& 1 & RMSprop & 0.03 & 0.999 & - & 0.02 & 0.92 & 10000 & 20 & 10000\\
\Xcline{1-13}{0.5pt}
G38 & 0.4 & 8e-4& 1 & RMSprop & 0.3 & 0.9999 & - & 0.0113 & 0.8595 & 7000 & 260 & 1000\\
\Xcline{1-13}{0.5pt}
G39  & 0.76 & 1.5e-4& 1 & RMSprop & 0.064 & 0.9999 & - & 0.0264 & 0.9081 & 7000 & 200 & 1000\\
\Xcline{1-13}{0.5pt}
G40 & 0.95 & 1.1e-4& 1 & RMSprop & 0.0525 & 0.9999 & - & 0.029 & 0.9082 & 10000 & 150 & 1000\\
\Xcline{1-13}{0.5pt}
G41 & 0.655 & 1.32e-5& 4.61 & SGD & 1.345 & - & 0.0725 & 0.0092 & 0.897 & 12000 & 200 & 1000\\
\Xcline{1-13}{0.5pt}
G42 & 1 & 1e-4 & 1 & RMSprop & 0.096 & 0.9999 & - & 0.024 & 0.73275 & 8000 & 10 & 10000\\
\Xcline{1-13}{0.5pt}
G43 & 0.65 & 6e-4 & 1 & SGD & 6.29 & - & 0.077 & 0.0285 & 0.7515 & 1000 & 30 & 1000\\
\Xcline{1-13}{0.5pt}
G44 & 0.65 & 7e-4 & 1.2 & SGD & 5.8 & - & 0.097 & 0.026 & 0.7554 & 1000 & 30 & 1000\\
\Xcline{1-13}{0.5pt}
G45 & 0.63 & 8.4e-4 & 1.36 & SGD & 6.1 & - & 0.129 & 0.01 & 0.755 & 3000 & 70 & 1000\\
\Xcline{1-13}{0.5pt}
G46 & 0.504 & 8.11e-6 & 2.07 & SGD & 1.54 & - & 0.156 & 0.0295 & 0.8965 & 2000 & 120 & 1000\\
\Xcline{1-13}{0.5pt}
G47 & 0.58 & 5.4e-4 & 1 & SGD & 7.5 & - & 0.13 & 0.026 & 0.76 & 3000 & 70 & 1000\\
\Xcline{1-13}{0.5pt}
G48 & 1.34 & 1e-3 & 1 & SGD & 5.5 & - & 0.08 & 0.032 & 0.737 & 180 & 3 & 1000\\
\Xcline{1-13}{0.5pt}
G49 & 1.77 & 5.9e-4 & 1 & SGD & 6.415 & - & 0.42 & 0.073 & 0.572 & 200 & 6 & 1000\\
\Xcline{1-13}{0.5pt}
G50 & 22.94 & 3.54e-5 & 0.833 & SGD & 0.436 & - & 0.0617 & 0.0503 & 0.3335 & 200 & 10 & 1000\\
\Xcline{1-13}{0.5pt}
G51 & 1.48 & 6.5e-6 & 1 & SGD & 1.345 & - & 0.283 & 0.029 & 0.863 & 7000 & 200 & 1000\\
\Xcline{1-13}{0.5pt}
G52 & 0.604 &4.2e-6 & 2.4 & SGD &2.9 & - & 0.19 & 0.027 & 0.81 & 10000 & 250 & 1000\\
\Xcline{1-13}{0.5pt}
G53 &0.27 &3.5e-6 & 10.8 & SGD &6 & - & 0.35 & 0.015 & 0.79 & 10000 & 250 & 1000\\
\Xcline{1-13}{0.5pt}
G54 &0.63 &1e-5 & 6 & SGD &1.27 & - & 0.11 & 0.018 & 0.71 & 10000 & 20 & 10000\\

\end{longtable}
}}

\subsubsection{Effects of the number of replicas to the cut-value distribution}
\begin{figure}[ht]
    \centering
    \includegraphics[width=1\linewidth]{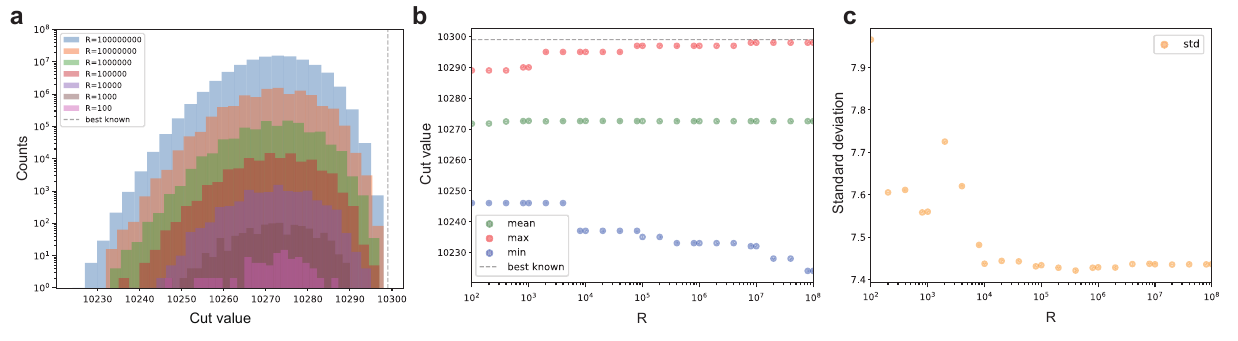}
    \caption{\textbf{The cut values as a function of the number of replicas $R$ used in FEM for the MaxCut Problem on the G55 dataset.}  \textbf{(a)}. The histograms of the cut values with the different $R$. \textbf{(b)}. The maximum, minimum, and mean cut values as a function of $R$. \textbf{(c)}. The standard deviation of the cut-value distribution in replicas with $R$ changes. }
    \label{replica_test}
\end{figure}

We explored how varying the number of replicas impacts the cut value distribution among replicas in the MaxCut problem on G55 in the G-set dataset. It is a random graph with 5000 nodes and $\{+1,-1\}$ edge weights. After optimizing FEM's hyperparameters, we incrementally increased the number of replicas $R$ to examine changes in the cut value distribution. The histograms of the cut values with different $R$ values are shown in Fig.~\ref{replica_test}(a). In Fig.~\ref{replica_test}(b), we found that the  average cut value remains stable with different $R$ values crossing several magnitudes. We also see that the standard deviation is also quite stable as shown in Fig.~\ref{replica_test}(c). As a consequence, the maximum cut value achieved by FEM is an increasing function of $R$.  It is clearly shown in Fig.~\ref{replica_test}(a) that the maximum cut value of FEM approaches the best-known results for G55 when $R$ increases. These findings also suggest that the hyperparameters of FEM can be fine-tuned using the mean cut value at a small $R$, while the final results can be obtained using a large R with fine-tuned parameters.


\subsection{Additional numerical experiments on the balanced minimum cut problem}
\subsubsection{Demonstration of FEM solving the bMinCut problem}
\begin{figure}[htbp]
\centering
\includegraphics[width=1\linewidth]{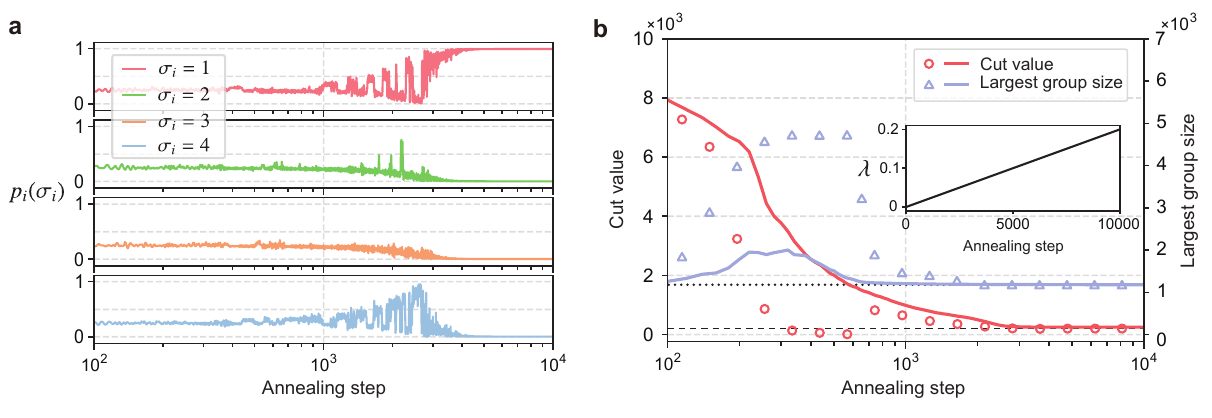}
\caption{\textbf{Illustrative example of the 4-way perfectly bMinCut problem on the real-world graph \emph{3elt}.}  \textbf{(a)}. The evolutions of the state probabilities of four typical spins in four allowed state classes. \textbf{(b)}. The evolutions of the cut values and the largest class sizes in 1000 replicas. 
 The circles denote the minimum cut values, and triangles indicate the maximum values of the largest class size, with solid lines illustrating the average values. The dotted line indicates the perfectly balanced class size 1180. The dashed line indicate the best known minimal cut 201. Inset: The hyperparameter $\lambda$ is linearly increased with the annealing step from 0 to $\lambda_{\rm max}=0.2$.}
\label{fig:bmincut}
\end{figure}

To elucidate the principles of FEM in tackling $q$-way bMinCut problem in the main text, firstly, we provide an example to demonstrate the numerical details in our experiments. We employ the real-world graph named \emph{3elt} as a demonstrative case study. This graph, which is collected in Chris Walshaw's graph partitioning archive~\cite{walshawgraph}, comprises 4,720 nodes and 13,722 edges, and we specifically address its 4-way bMinCut problem as an illustrative example. 
Fig.~\ref{fig:bmincut}(a) shows the evolution of the marginal probabilities for $4$ states for a typical variable $\sigma_i$. From the figure, we can identify $3$ optimization stages. At early annealing steps, the $4$ marginal probabilities $\{P_i(\sigma_i=1),P_i(\sigma_i=2),P_i(\sigma_i=3),P_i(\sigma_i=4)\}$ are all very close to the initial value $0.25$; with annealing step increases, the marginal probabilities for $4$ states begin to fluctuate; then finally converge. Only one probability converges to unity and the other three probabilities converge to $0$. This indicates that during annealing and the minimization of mean-field free energy, the marginal probability will evolve towards localizing on a single state. In Fig.~\ref{fig:bmincut}(b), we plot the evolutions of the cut values and the largest group sizes (the largest value of the sizes among all groups) averaged over $R=1000$ replicas of mean-field approximations. 
In our approach, the hyperparameter $\lambda$ linearly increases from 0 to $\lambda_{\max}$ with the annealing step, for preferentially searching the states that minimize the cut value at the initial optimization stage. From the figure, we can see that our algorithm first searches for a low-cut but un-balanced solution with a large group and three small groups, then gradually decreases the maximum group, and finally finds a balanced solution with a global minimum cut.

\subsubsection{Benchmarking experiments for the bMinCut problem on the real-world graphs}
In our benchmarking experiments, we utilized the latest release of the METIS software, with version 5.1.0~\cite{metis5.1.0}, specifically employing the stand-alone program named \emph{gpmetis} for the bMinCut tasks. Throughout the experiment, we consistently configured \emph{gpmetis} with specific options as outlined in the documentation~\cite{metis5.1.0}: \emph{-ptype=rb}, which denotes multilevel recursive bisectioning, and \emph{-ufactor=1} (1 is the lowest permissible integer value allowed by METIS, indicating that while a perfectly balanced partitioning cannot be guaranteed, we aim to approach it as closely as possible). All other parameters or options were left at their default settings.

For KaHIP, we used the advanced variant of KaHIP named KaFFPaE~\cite{kahipgithub} for the bMinCut tasks, with the following settings (see also the documentation~\cite{sanders2013kahip} for more explanations): \emph{-n=24} (number of processes to use), \emph{--time\_limit=300} (limiting the execution time to 300 seconds), \emph{--imbalance=0}, \emph{--preconfiguration=strong}, and we also enabled \emph{--mh\_enable\_tabu\_search}  and \emph{--mh\_enable\_kabapE} to optimize performance. Both the programs of \emph{gpmetis} and KaFFPaE, implemented in C, were executed on 
a computing node equipped with dual 24-core processors with 2.90GHz. 

For FEM, we have detailed  hyperparameter settings in Tab.~\ref{tab:bmincut_param}. Throughout the experiments, we initialize the local fields with random values according to $h_i(\sigma_i)^{ini} = 0.001 * {\rm randn}$, where ${\rm randn}$ represents a random number sampled from the standard Gaussian distribution. The execution of FEM was performed on a GPU. All variables are represented using 32-bit single-precision floating-point numbers.

{\renewcommand{\arraystretch}{1.5}
\setlength{\tabcolsep}{1mm}
\setlength{\LTcapwidth}{\textwidth}
{\footnotesize
\begin{longtable}[c]{ccccccccccccc}
\caption{\textbf{The hyperparameter settings for FEM in the bMinCut benchmarking on real-world graphs.} The optimization algorithm employed in this experiment is Adam. Note, the imbalance penalty coefficient, $\lambda$, is progressively incremented from 0 to its maximum value, $\lambda_{max}$, over $N_{\rm step}$ steps. The term RC time refers to the abbreviation of replica computation time when running FEM. The exponential scheduling is used for the annealing. \label{tab:bmincut_param}}\\ 
\Xcline{1-13}{0.9pt}
\multirow{2}*{Graph} & \multirow{2}*{\textit{\textbf{q}}} &\multirow{2}*{\boldmath{$\beta_{min}$}}&\multirow{2}*{\boldmath{$\beta_{max}$}}&\multirow{2}*{\boldmath{$\lambda_{max}$}}&\multirow{2}*{\boldmath{$N_{\rm step}$}}&\multirow{2}*{\boldmath{$N_{\rm replica}$}}&\multirow{2}*{\boldmath{$\gamma_{\rm grad}$}}&\multirow{2}*{\textit{\textbf{lr}}}&\multirow{2}*{\makecell[c]{\textbf{weight}\\\textbf{decay}}}&\multirow{2}*{\boldmath{$\beta_1$}}&\multirow{2}*{\boldmath{$\beta_2$}}&\multirow{2}*{\makecell[c]{\textbf{RC}\\\textbf{time(sec.)}}}\\
~&~&~&~&~&~&~&~&~&~&~&~&~\\
\Xcline{1-13}{0.9pt}
\endfirsthead

\Xcline{1-13}{0.9pt}
\multirow{2}*{Graph} & \multirow{2}*{\textit{\textbf{q}}} &\multirow{2}*{\boldmath{$\beta_{min}$}}&\multirow{2}*{\boldmath{$\beta_{max}$}}&\multirow{2}*{\boldmath{$\lambda_{max}$}}&\multirow{2}*{\boldmath{$N_{\rm step}$}}&\multirow{2}*{\boldmath{$N_{\rm replica}$}}&\multirow{2}*{\boldmath{$\gamma_{\rm grad}$}}&\multirow{2}*{\textit{\textbf{lr}}}&\multirow{2}*{\makecell[c]{\textbf{weight}\\\textbf{decay}}}&\multirow{2}*{\boldmath{$\beta_1$}}&\multirow{2}*{\boldmath{$\beta_2$}}&\multirow{2}*{\makecell[c]{\textbf{RC}\\\textbf{time(sec.)}}}\\
~&~&~&~&~&~&~&~&~&~&~&~&~\\
\Xcline{1-13}{0.9pt}
\endhead
\multicolumn{13}{c}{\textbf{\normalsize to be continued...}}
\endfoot
\Xcline{1-13}{0.9pt}
\endlastfoot

\multirow{5}*{\emph{add20}} & 2  & 2.12e-2 & 756 & 1 & 10000 &2000&17.8&0.2914&3.4e-4&0.9408&0.7829& 24.2\\
\Xcline{2-13}{0.5pt}
~ & 4  & 5.054e-2 & 840.84 & 0.2332 &10000 &2000&13.33&0.3664&3.411e-3&0.9158&0.7691&46.2\\
\Xcline{2-13}{0.5pt}
~ & 8& 3.048e-2 & 2178.64 & 0.6229 & 10000 &2000&10.05&0.4564&4.198e-4&0.9018&0.7225&90.2\\
\Xcline{2-13}{0.5pt}
~ & 16 & 3.634e-2 & 1607.45 & 1.0553 & 10000 &2000&81.04&0.6564&8.264e-3&0.9032&0.6009&181.2\\
\Xcline{2-13}{0.5pt}
~ & 32 & 3.77e-2 & 2827.44 & 1.9607 & 10000 & 2000 &7.384&1.4246&0.01719&0.911&0.8199&360.1\\
\Xcline{1-13}{0.8pt}
\multirow{5}*{\emph{data}} & 2 & 5.054e-2 & 840.84 & 0.2292 & 10000 &2000&13.328&0.2629&1.706e-3&0.9347&0.7692&29.1\\
\Xcline{2-13}{0.5pt}
~ & 4 &  5.054e-2 & 840.84 & 0.2292 & 10000 &2000&13.328&0.2629&1.706e-3&0.9347&0.7692&57.1\\
\Xcline{2-13}{0.5pt}
~ & 8 & 5.054e-2 & 840.84& 0.3438 & 10000 &2000&13.328&0.3023&1.365e-3&0.9369&0.8076&109.2\\
\Xcline{2-13}{0.5pt}
~ & 16  & 7.316e-2 & 1766.14 & 1.3376 & 10000 &2000&15.736&0.8358&1.874e-4&0.7801&0.4039&217.5\\
\Xcline{2-13}{0.5pt}
~ & 32  & 1.968e-2 & 613.88 & 0.697 & 12000 &2000&57.56&0.6633&1.65e-3&0.5263&0.4681&519\\
\Xcline{1-13}{0.8pt}
\multirow{5}*{\emph{3elt}} & 2 & 3.648e-2 & 2458.64 & 2.07 & 10000 &2000&7.146&1.2032&2.293e-2&0.9374&0.9215&45.8\\
\Xcline{2-13}{0.5pt}
~ & 4  & 3.648e-2 & 2458.64 & 2.07 & 10000 &2000&7.146&1.2032&2.293e-2&0.9374&0.9215&90.2\\
\Xcline{2-13}{0.5pt}
~ & 8  & 2.918e-2 & 1966.92 & 0.8 & 10000 &2000&7.384&0.2117&1.031e-3&0.724&0.5397&177.1\\
\Xcline{2-13}{0.5pt}
~ & 16  & 3.648e-2 & 2458.64 & 1.267 & 10000 &2000&7.146&0.2238&2.577e-3&0.77&0.7698&355.3\\
\Xcline{2-13}{0.5pt}
~ & 32  & 4.626e-2 & 3581.42 & 1.369 & 12000 &2000&6.768&0.9946&2.508e-2&0.777&0.8982&850\\
\Xcline{1-13}{0.8pt}
\multirow{5}*{\emph{bcsstk33}} & 2 & 3.648e-2 & 2458.64 & 2.07 & 10000 &2000&7.146&1.2032&2.2926e-2&0.9374&0.9215&103\\
\Xcline{2-13}{0.5pt}
~ & 4 & 3.648e-2 & 2458.64 & 2.07  & 10000 &2000&7.146&1.2032&2.2926e-2&0.9374&0.9215&204\\
\Xcline{2-13}{0.5pt}
~ & 8  & 3.72e-2 & 754.22 & 0.5016 & 12000 &2000&15.6&0.1948&1.663e-3&0.7196&0.9&480\\
\Xcline{2-13}{0.5pt}
~ & 16  & 3.336e-2 & 950 & 1.167 & 12000 &2000&7.052&0.5254&3.089e-3&0.8894&0.6223&968\\
\Xcline{2-13}{0.5pt}
~ & 32 & 2.918e-2 &2827.44 & 1.656 & 12000 &2000&7.384&1.5541&1.146e-2&0.9582&0.8686&1986\\
\end{longtable}
}
}

\subsubsection{More details on deploying large-scale operators in the chip verification task}

\begin{figure}[t]
\centering
\includegraphics[width=0.9\linewidth]{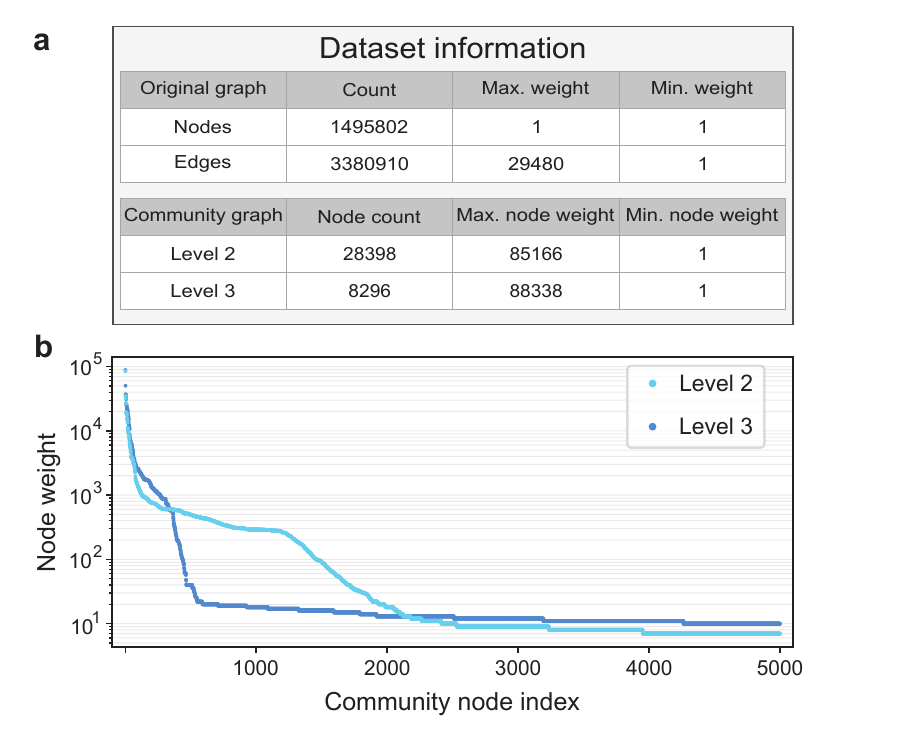}
\caption{\textbf{The detailed information of the large-scale realistic dataset used in this chip verification task.}  \textbf{(a)}. The information of the original graph and the two community graphs given by the Louvain algorithm. \textbf{(b)}. The top 5000 community nodes with the largest node weight in the two community graphs.}
\label{fig:fpga_data}
\end{figure}

In this chip verification task described in the main text, the large-scale realistic dataset consists of 1495802 operators and 3380910 logical operator connections. We map this dataset into an undirected weighted graph with 1495802 nodes and 3380910 edges. The information of this original graph is listed in Fig.~\ref{fig:fpga_data}(a). Note that, in the original graph, there are many edges with weights larger than 1, which differs from the other graphs used in bMinCut benchmarking.

Besides, since the realistic datasets in this task often have locally connected structures among operators, meaning many nodes form local clusters or community structures, we consider applying the coarsening trick to reduce the graph size for the sake of performance and speed. A good option is to coarsen the graph based on its natural community structure. We group the nodes within the same community into a new community node, hierarchically contracting the original graph. We refer to the coarsened graph as the community graph. Note that in the community graph, the community nodes now have a node weight larger than 1, equal to the number of nodes grouped in the community. The weight of the community edges connecting different community nodes is thus the sum of the original edges that have endpoints in different communities.

To effectively identify the community structure in a large-scale graph, we apply the Louvain algorithm~\cite{blondel2008fast}, which was developed for community detection. The information of the community graphs output from the Louvain algorithm, corresponding to level 2 and level 3, is also listed in Fig.~\ref{fig:fpga_data}(a). The modularity metric of the original graph is 0.961652. Fig.~\ref{fig:fpga_data}(b) shows the top 5000 community nodes with the largest node weight in the two community graphs. Note that the coarsening technique used to generate coarsened graphs here is different from the methods based on maximal matching techniques employed in METIS.

Since FEM now has to address the issue of ensuring the balanced constraint on nodes with different node weights, a slight modification on the constraint term in the formula of mean-field free energy should be made, which is 
\begin{align}
    \lambda\sum_{i,j}\sum_{\sigma_i}[P_i(\sigma_i)P_j(\sigma_i)-P^2_i(\sigma_i)] \rightarrow \lambda\sum_{i,j}\sum_{\sigma_i}[V_iV_jP_i(\sigma_i)P_j(\sigma_i)-V_i^2P^2_i(\sigma_i)]  \;,
\end{align}
where $V_i$ is the node weight of the $i$-th node. The corresponding modifications on gradients are 
\begin{align}
g^p_i(\sigma_i): 2\lambda[\sum_jP_j(\sigma_i)-P_i(\sigma_i)] \rightarrow 2\lambda V_i[\sum_jV_jP_j(\sigma_i)-V_iP_i(\sigma_i)]\;,
\end{align}
\begin{gather}
\hat{g}_i^p(\sigma_i) : \lambda c_i^a[\sum_je_j(\sigma_i)-e_i(\sigma_i)] \rightarrow \lambda c_i^aV_i[\sum_jV_je_j(\sigma_i)-V_ie_i(\sigma_i)]\; ,
\end{gather}
while keeping other things unchanged.

For better performance, we choose the community graph for partitioning according to the different values of $q$. For $q=4,8$, we use the graph ``level 3". For $q=16$, we use the graph ``level 2". The main reason is that the larger the value of $q$, the smaller the largest group size due to the balanced constraint. Therefore, we have to pick the community graph with a smaller maximum node weight to give enough room for the search. If the community graph has a maximum node weight larger than the allowed largest group size, then the balanced constraint condition with respect to the original graph will always be violated.

\subsection{Numerical experiments on the Max $k$-SAT problem}
For the benchmarking experiments targeting the Max $k$-SAT problem, we collated the reported data encompassing the best-known results and corresponding shortest computation times from the most efficient incomplete solvers. These solvers were evaluated against 454 random instances from the random MaxSAT 2016 competition problems~\cite{maxksat2016}, with the summarized data presented in Tab.~\ref{tab:maxksat1}. The incomplete solvers included are SsMonteCarlo, Ramp, borealis, SC2016, CCLS, CnC-LS, Swcca-ms, HS-Greedy, and CCEHC. For the FEM approach, we have also compiled the summarized results in Tab.~\ref{tab:maxksat1} and the hyperparameter settings in Tab.~\ref{tab:maxksat2}. Still, throughout the experiments, we initialize the local fields with random values according to $h_i(\sigma_i)^{ini} = 0.001 * {\rm randn}$, where ${\rm randn}$ represents a random number sampled from the standard Gaussian distribution. The implementation of FEM was executed on a GPU to ensure efficient computation. All variables are represented using 32-bit single-precision floating-point numbers.

{\renewcommand{\arraystretch}{1.5}
\setlength{\tabcolsep}{0.6mm}
\setlength{\LTcapwidth}{0.9\textwidth}
{\footnotesize
\begin{longtable}[c]{cccccccc}
\caption{\textbf{The results for FEM in the benchmarking on the random MaxSAT 2016 competition problems.} RC time refers to the abbreviation of replica computation time, and $\Delta_E = E_{\rm min}-E_{\rm bkr}$. $E_{\rm min}$ is the minimal energy found by FEM, and $E_{\rm bkr}$ is the best known value in the literature. \label{tab:maxksat1}}\\
\Xcline{1-8}{0.9pt}
\multirow{2}*{\textbf{Index}} & \multirow{2}*{\textbf{Instance}} &\multirow{2}*{\boldmath{$E_{\rm bkr}$}}&\multirow{2}*{\makecell[c]{\textbf{Fastest}\\\textbf{solver}}}&\multirow{2}*{\makecell[c]{\textbf{Solver's}\\\textbf{time(sec.)}}}&\multirow{2}*{\boldmath{$E_{\rm min}$}}&\multirow{2}*{\boldmath{$\Delta_E$}}&\multirow{2}*{\makecell[c]{\textbf{RC}\\\textbf{time(sec.)}}}\\
~&~&~&~&~&~&~&~\\
\Xcline{1-8}{0.9pt}
\endfirsthead

\Xcline{1-8}{0.9pt}
\multirow{2}*{\textbf{Index}} & \multirow{2}*{\textbf{Instance}} &\multirow{2}*{\boldmath{$E_{\rm bkr}$}}&\multirow{2}*{\makecell[c]{\textbf{Fastest}\\\textbf{solver}}}&\multirow{2}*{\makecell[c]{\textbf{Solver's}\\\textbf{time(sec.)}}}&\multirow{2}*{\boldmath{$E_{\rm min}$}}&\multirow{2}*{\boldmath{$\Delta_E$}}&\multirow{2}*{\makecell[c]{\textbf{RC}\\\textbf{time(sec.)}}}\\
~&~&~&~&~&~&~&~\\
\Xcline{1-8}{0.9pt}
\endhead
\Xcline{1-8}{0.5pt}
\multicolumn{8}{c}{\textbf{\normalsize to be continued...}}
\endfoot
\Xcline{1-8}{0.9pt}
\endlastfoot
1&s2v120c2600-2&458&SsMonteCarlo&1.75&458&0&0.055\\
\Xcline{1-8}{0.5pt}
2&s2v120c2600-3&440&SsMonteCarlo&1.34&440&0&0.055\\
\Xcline{1-8}{0.5pt}
3&s2v120c2600-1&439&SsMonteCarlo&1.41&439&0&0.053\\
\Xcline{1-8}{0.5pt}
4&s2v120c2500-2&435&Ramp&1.36&435&0&0.055\\
\Xcline{1-8}{0.5pt}
5&s2v120c2500-3&425&SsMonteCarlo&1.62&425&0&0.056\\
\Xcline{1-8}{0.5pt}
6&s2v140c2600-2&425&Ramp&2.0&425&0&0.049\\
\Xcline{1-8}{0.5pt}
7&s2v140c2600-1&422&Ramp&2.12&422&0&0.056\\
\Xcline{1-8}{0.5pt}
8&s2v120c2500-1&418&SsMonteCarlo&1.53&418&0&0.056\\
\end{longtable}}
}

{\renewcommand{\arraystretch}{1.5}
\setlength{\tabcolsep}{0.6mm}
\setlength{\LTcapwidth}{0.9\textwidth}
{\footnotesize
\begin{longtable}[c]{ccccccccccc}
\caption{\textbf{The hyperparameter settings for FEM in the benchmarking on the random MaxSAT 2016 competition problems.} Throughout the experiments, we employed the RMSprop as the optimizer. The inverse-proportional scheduling is used for the annealing. \label{tab:maxksat2}}\\
\Xcline{1-11}{0.9pt}

\multirow{2}*{\textbf{Index}} & \multirow{2}*{\textbf{Instance}} &\multirow{2}*{\boldmath{$N_{\rm step}$}}&\multirow{2}*{\boldmath{$N_{\rm replica}$}}&\multirow{2}*{\boldmath{$T_{\rm min}$}}&\multirow{2}*{\boldmath{$T_{\rm max}$}}&\multirow{2}*{\boldmath{$\gamma_{\rm grad}$}}&\multirow{2}*{\textit{\textbf{lr}}}&\multirow{2}*{\textbf{alpha}}&\multirow{2}*{\makecell[c]{\textbf{weight}\\\textbf{decay}}}&\multirow{2}*{\makecell[c]{\textbf{momen-}\\\textbf{tum}}}\\
~&~&~&~&~&~&~&~&~&~&~\\
\Xcline{1-11}{0.9pt}
\endfirsthead
\Xcline{1-11}{0.9pt}
\multirow{2}*{\textbf{Index}} & \multirow{2}*{\textbf{Instance}} &\multirow{2}*{\boldmath{$N_{\rm step}$}}&\multirow{2}*{\boldmath{$N_{\rm replica}$}}&\multirow{2}*{\boldmath{$T_{\rm min}$}}&\multirow{2}*{\boldmath{$T_{\rm max}$}}&\multirow{2}*{\boldmath{$\gamma_{\rm grad}$}}&\multirow{2}*{\textit{\textbf{lr}}}&\multirow{2}*{\textbf{alpha}}&\multirow{2}*{\makecell[c]{\textbf{weight}\\\textbf{decay}}}&\multirow{2}*{\makecell[c]{\textbf{momen-}\\\textbf{tum}}}\\
~&~&~&~&~&~&~&~&~&~&~\\
\Xcline{1-11}{0.9pt}
\endhead
\multicolumn{11}{c}{\textbf{\normalsize to be continued...}}
\endfoot
\Xcline{1-11}{0.9pt}
\endlastfoot
1&s2v120c2600-2&50&50&0.0001&0.6&0.8&0.8&0.3&0.02&0.9\\
\Xcline{1-11}{0.5pt}
2&s2v120c2600-3&50&50&0.0001&0.6&0.8&0.8&0.3&0.02&0.9\\
\Xcline{1-11}{0.5pt}
3&s2v120c2600-1&50&50&0.0001&0.6&0.8&0.8&0.3&0.02&0.9\\
\Xcline{1-11}{0.5pt}
4&s2v120c2500-2&50&50&0.0001&0.6&0.8&0.8&0.3&0.02&0.9\\
\Xcline{1-11}{0.5pt}
5&s2v120c2500-3&50&50&0.0001&0.6&0.8&0.8&0.3&0.02&0.9\\
\Xcline{1-11}{0.5pt}
6&s2v140c2600-2&50&50&0.0001&0.6&0.8&0.8&0.3&0.02&0.9\\
\Xcline{1-11}{0.5pt}
7&s2v140c2600-1&50&50&0.0001&0.6&0.8&0.8&0.3&0.02&0.9\\
\Xcline{1-11}{0.5pt}
8&s2v120c2500-1&50&50&0.0001&0.6&0.8&0.8&0.3&0.02&0.9\\
\end{longtable}}
}

\subsection{The mean field equations for the Ising problem derived from the FEM formalism}
We introduce reproducing the mean-field equations for the Ising problem from the FEM formalism. For the case of Ising problem with the Ising Hamiltonian $E_{Ising}(\bm{\sigma})= -\frac{1}{2}\sum_{i,j}^NW_{ij}\sigma_i\sigma_j$, where $\sigma_i \in \{-1,+1\}$, the corresponding mean field free energy $F_{\rm MF} = U_{\rm MF}-\frac{1}{\beta}S_{\rm MF}$ is computed as follows. The the internal energy $U_{\rm MF}$ is defined as the expectation of $E_{Ising}(\bm{\sigma})$ regarding to mean field joint probability $P_{\rm MF}(\bm{\sigma})=\prod_i^NP_i(\sigma_i)$, which reads 

\begin{align}   
U_{\rm MF} &= -\frac{1}{2}\sum_{\bm{\sigma}}\prod_iP_i(\sigma_i)\sum_{i,j}W_{ij}\sigma_i\sigma_j\nonumber\\
&=-\frac{1}{2}\sum_{i,j}W_{ij}\left[\sum_{\sigma_i}P_i(\sigma_i)\sigma_i\right]\left[\sum_{\sigma_j}P_j(\sigma_j)\sigma_j\right]\nonumber\\
& = -\frac{1}{2}\sum_{i,j}W_{ij} m_im_j \; ,
\end{align}
where $m_i=\sum_{\sigma_i}P_i(\sigma_i)\sigma_i=P_i(+1)-P_i(-1)$ is the magnetization (or mean spin). And the entropy $S_{\rm MF}$ reads
\begin{align}   
S_{\rm MF} & =-\sum_i\sum_{\sigma_i}P_i(\sigma_i) \ln P_i(\sigma_i)\nonumber\\
&=-\sum_{i}\sum_{\sigma_i}\frac{1+m_i\sigma_i}{2}\ {\rm ln}\ \frac{1+m_i\sigma_i}{2}\nonumber\\
&=-\sum_{i}\left(\frac{1+m_i}{2}\ {\rm ln}\ \frac{1+m_i}{2} + \frac{1-m_i}{2}\ {\rm ln}\ \frac{1-m_i}{2}\right) \; ,
\end{align}
where the identity relation $P_i(\sigma_i)=\frac{1+m_i\sigma_i}{2}$ is used.
Accordingly, the mean field free energy for the Ising Hamiltonian turns out to be
\begin{align}   
\label{Seq:mfF_wst_m}
F_{\rm MF} =  -\frac{1}{2}\sum_{i,j}W_{ij} m_im_j + \frac{1}{\beta}\sum_i\left(\frac{1+m_i}{2}\ {\rm ln}\ \frac{1+m_i}{2} + \frac{1-m_i}{2}\ {\rm ln}\ \frac{1-m_i}{2}\right) \; .
\end{align}

At any given $\beta$, the optimal $\{P^*_i(\sigma_i)\}$ that minimizes $F_{\rm MF}$ can be obtained by the zero-gradient equations $\frac{\partial F_{\rm MF}}{\partial P_i(+1)}=-\frac{\partial F_{\rm MF}}{\partial P_i(-1)}=0$ (since $P_i(+1)+P_i(-1)=1$). And since $P_i(\sigma_i)=\frac{1+m_i\sigma_i}{2}$, it indicates that the information of all marginal probabilities is completely described by the magnetizations. Thus, it suffices to solve the zero-gradient equations with respect to $\{m_i\}$, which is
\begin{align}
\frac{\partial F_{\rm MF}}{\partial m_i}=0
 &=  -\sum_{j}W_{ij}m_j + \frac{1}{\beta}\left(\frac{1}{2}{\rm ln}\frac{1+m_i}{1-m_i}\right)\nonumber\\
 &= -\sum_{j}W_{ij}m_j + \frac{1}{\beta}{\rm arctanh}(m_i) \; .
\end{align}
As a result, we find that 
\begin{align}
\label{Seq:m_iterate}
m_i ={\rm tanh}\,\Bigg(\beta\sum_{j}W_{ij}m_j\Bigg),\ \ \ i,j= 1, 2, ..., N \; ,
\end{align}
which reproduces the mean field equations in the existing mean-field annealing approaches for combinatorial optimization. Then, solving Eq.~(\ref{Seq:m_iterate}) by using the fixed-point iteration method with a slow annealing for $\beta\rightarrow \infty$ leads to the solution $\{m^*_i\}$ to Eq.~(\ref{Seq:m_iterate}). We can identify the ground state by the operation of $\sigma^{GS}_i=\arg \max_{\sigma_i}P^*_i(\sigma_i)={\rm sign}(m^*_i)$.

\subsection{Modelings for other combinatorial optimization problems\label{sec:otherCOPs}}
Building upon the three example problems explored in the main text, we now outline a concise methodology for modeling or formulating additional well-known COPs within the FEM framework. As demonstrated previously, the essential step for solving a given COP using FEM is to precisely define the expectation of the designed cost function with respect to the marginal probabilities. Given the fixed form of the entropy in the mean-field ansatz, calculating the internal energy provides the complete free energy expression that is a function of the marginals. Subsequently, the COP can be tackled using either the automatic differentiation approach or the explicit gradient formulations.

\subsubsection{COPs that can be naturally translated into QUBO formulations}
Probably the most simple and straightforward ones will be problems that can be naturally translated into QUBO problems (or Ising formulations).
The target function of QUBO problems can be written as
\begin{equation}
    E(\bm{\sigma}) = \bm{\sigma}^TQ\bm{\sigma} = \sum_{i,j} Q_{ij}\sigma_i \sigma_j \;.
\end{equation}
Here $\bm{\sigma}$ is a vector of binary spin variables (0 or 1) and $Q$ is a symmetric matrix encoding the QUBO problems.

The expectation value of the cost function, or $U_{\rm MF}$, will be
\begin{equation}
    \langle E \rangle = \sum_i\sum_j p_i Q_{ij} p_j \;,
\end{equation}
with the $p_i$ represents the marginal probability of spin variable $\sigma_i$ to take on value $1$.

Many kinds of COPs are either naturally QUBO or can be translated into QUBO problems with simple steps. Some prominent such examples are listed below.
\begin{enumerate}
    \item Vertex cover
    \item Maximum Independent Set Problems
    \item Binary Integer Linear Programming
    \item Set Packing
    \item Maximum cliques
\end{enumerate}

\subsubsection{Problems of multi-valued variables}
The COPs with multi-state variables, commonly referred as Potts model in the statistical physics community, have different target function formulation with the aforementioned QUBO-like problems. In the main text, we have shown a example of balanced minimum cut problem. Another such problem will be the graph coloring problems, which detects whether a coloring configuration of a graph can satisfy that no end points of an edge have the same color. This problem can be transferred into a binary variable case with ancillary extra variables~\cite{lucas2014ising}, the cost function then will be
\begin{equation}
    E(\sigma) = \sum_i\left(1-\sum_c\sigma_{i, c}\right)^2 + 
\sum_{(i,j)} \sum_c \sigma_{i,c} \sigma_{j,c} \; .
\end{equation}
Here there are overall $N\times C$ binary variables $\sigma_{i,c}$ with $N$ being the number of vertices, $C$ representing the overall number of colors and $\sigma_{i,c}$ being a variable to determine whether vertex $i$ has the color $c$. The first term of this cost function is a regularization term to let all vertices to have only one color and the second term is the actual target function which describes the number of violation edges whose two end vertices have the same color.

If we change the framework into the FEM solver, modeling will become much easier with only $N$ multi-state variables and no regularization term. The target function can be written as 
\begin{equation}
    E(\bm{\sigma}) = \sum_{(i,j)} \delta(\sigma_i, \sigma_j) \; ,
\end{equation}
with $\sigma_i$ being a multi-valued spin variable which represents the color of vertex $i$.
The expectation value of such target function can be deducted easily as 
\begin{equation}
    \langle E(\bm{\sigma}) \rangle = \sum_{(i,j)}\sum_{\sigma_i} p_i(\sigma_i) p_j(\sigma_i) \; .
\end{equation}

With the graph coloring problem being the natural one, there are many other problems can be translated into such formulations. We have listed some of the examples below.
\begin{itemize}
    \item Hamiltonian Cycles and Paths
    \item Traveling Salesman
    \item Community Detection
\end{itemize}

\subsubsection{Problems with multi-variable interactions}
For problems like SAT and quantum error corrections (QEC), the target function will encounter terms with multi-variable interactions. In the main text, we have already introduce how to solve Max $k$-SAT. Here we demonstrate the formulation of QEC that compatible to FEM solver.

The QEC problems can be translated into Ising spin-glass problems with the Hamiltonian of the variable $\sigma_i\in\{-1, +1\}$~\cite{fujisaki2022Practical,fujisaki2023Quantum}
\begin{equation}
    H = -J\sum_v^{N_v}b_v\prod_{i\in \delta v}^4\sigma_i - h \sum_i^{N_d} \sigma_i \; ,
\end{equation}

The expectation value of such Hamiltonian can then be written as 
\begin{align}
    \langle E \rangle &= -J\sum_v^{N_v}b_v\prod_{i\in \delta v}^4(p_i(+1) - p_i(-1)) - h \sum_i^{N_d} (p_i(+1) - p_i(-1)) \nonumber \\
    &= -J\sum_v^{N_v}b_v\prod_{i\in \delta v}^4 m_i - h \sum_i^{N_d}m_i \; .
\end{align}
Here $m_i = p_i(+1) - p_i(-1)$ can be viewed as the magnetization of variable $i$.
\end{document}